\documentclass[a4paper,10pt]{article}

\title{ Cosmological perturbation theory to second order for curvature, density and
gravity waves on FRW background; and the WMAP results of inhomogeneity and clustering in the early
universe. }
\author{Ajay Patwardhan, Kartik Prabhu, M.S.R.Kumar}

\usepackage{subfig}
\usepackage{latexsym}

\usepackage{graphics}
\usepackage{graphicx}
\usepackage{geometry}
\begin{document}
\maketitle
\begin{abstract}

The second order perturbation calculations for gravity wave and
Einstein equation for space time and matter are presented for the
FRW metric cosmological model. While exact equations are found,
suitable approximations are made to obtain definite results. In the
gravity wave case the small wavelength case allows nearly locally
flat background  for obtaining a fit to the WMAP data. In the
density and curvature case the FRW background is retained for the
length scale of WMAP. Clustering and inhomogeneity are understood.
The gravity wave ripples from Big Bang couple nonlinearly and
redistribute the modes to higher values of 'l' giving consistency
with the WMAP results. The order by order consistency of Einstein
equations relate the second order perturbations in the curvature and
density and the wrinkles in spacetime caused by the gravity wave
modes reorganize these distributions. The radiation data of WMAP
gives the picture of a FRW spacetime deformed and wrinkled
consistent with matter distribution to one hundred thousandths parts
variation.

\end{abstract}
\section{INTRODUCTION}

The WMAP data at the observable horizon of $ 300000  $ years after
Big Bang gives confirmation of anisotropy and inhomogeneity in the
universe. A bubble model during inflation is made with composite
equation of state. An explanation of the anisotropy with WMAP $ C(l)
vs "l" $ is made with gravity wave mixing causing 'wrinkled' space
time from big bang to WMAP epoch. A density perturbation model
within a Newtonian and Einsteinian analysis is presented. The
clustering model attempts to explain the two fluid phase with
cluster formation at different length scales. The recent WMAP
picture shows that
           \(\frac{\triangle\rho}{\rho}\)and\(\frac{\triangle T}{T}\)
           of the order of \(10^{-5}\)\@.Then it evolves to clustering and
           clumping of structures\@.There are number of models for
           this clustering\@.In this project as a continuation of Part
           A and Part B\@.The density perturbations are evolved till 1
           billion years from Big Bang when galaxies are formed\@.Therefore we simulate the process of
           gravitational clustering as process of diffusion running
           backwards in time i.e, with an anti-diffusion
           model\@.A simple model of a uniform Gaussian distribution which can
           collapse to localized sharp distributions simulates the
           process of variation in density.We have constructed a 1-d
           Gaussian and a 2-d Gaussian depicting variation of
           density on a line and in a plane respectively\@.Based on
           this a 3-d Gaussian representing variation of density in
           volume may be understood\@.We are actually using Gaussian
           distributions as solutions of a diffusion equation with
           the following parameters the width of the Gaussian\@,the
           center of Gaussian and height of Gaussian\@.With these
           parameters a well spread Gaussian becomes narrower with
           time running backward and becomes wider with time
           running forward\@.Therefore we simulate the process of
           gravitational clustering as process of diffusion running
           backwards in time i.e, with an anti-diffusion
           model\@.This simulation is initially applied at the
           scale (4000 mega parsec) of the Universe leading to
           formation osuperclusters.Then within a supercluster we
           understand the process similarly to explain the formation of
           clusters.Finally inside a cluster we associate the
           formation of galaxies with a similar argument.

            The recent WMAP picture shows us that during its
            initial phase there was no differentiation between
            various forms of matter namely dark matter\@,dark
            energy\@,normal matter and radiation\@.But as the
            Universe evolved different forms of matter started
            differentiating among themselves and occupied
            different locations\@.To be more specific the
            redistribution of normal matter was more pronounced
            as large lumps of matter started accumulating
            together to form centers of galaxies which are surrounded by dark matter halos.
            Each halo is five times
            more massive
            than the normal matter at the
            center\@.Note that the normal matter at the center
            together with the dark matter halo forms the
            respective galaxy
            \@.Some of these galaxies
            came together to form clusters of galaxies and many
            such clusters approached each other to form super cluster of
            galaxies\@.The dark energy as well as dark matter occupied in
            the empty space which are called voids\@.The
            redistribution of dark energy is not as much as that
            of normal matter\@.Therefore an early Universe which
            had a uniform distribution of various forms of matter
            ended up with a one in which the various forms of
            matter got segregated at different places\@.The
            title of our report precisely calls for an answer to
            this observation\@.
\newline

                The overall properties of the Universe roughly speaking are close
                to being homogenous and yet telescopes reveal a wealth of detail on scales varying from single
                galaxies to large scale structures of size exceeding 100
                mega parsecs\@.The existence of these cosmological
                structures tells us something important about the
                initial conditions of the Big Bang,and about physical
                processes that have operated subsequently\@.If the initial
                Big Bang would have been homogenous then the presently
                observed distribution of galaxies cannot be accounted
                for considering the age of the Universe\@.If the
                elementary particles in the Universe started out with a
                uniform distribution\@,then purely statistical
                fluctuations would occur on all scales\@,from which
                matter would eventually condense\@.However the highly
                improbable fluctuations of galactic dimensions
                 is extremely slow and takes far more time
                compared to the age of the Universe\@.Therefore studying
                models with inhomogeneities is taken up\@.The aim of
                studying inhomogeneities is to understand the processes that caused the Universe to depart from
                uniform density\@
                The second order perturbation calculations for gravity wave and
Einstein equation for space time and matter are presented for the
FRW metric cosmological model. While exact equations are found,
suitable approximations are made to obtain definite results. In the
gravity wave case the small wavelength case allows nearly locally
flat background  for obtaining a fit to the WMAP data. In the
density and curvature case the FRW background is retained for the
length scale of WMAP. Clustering and inhomogeneity are understood.

\section{Bubble model}
\subsection*{Introduction}

    A model of the expansion of the Universe by the expansion of a bubble in some surrounding medium when
     the contents of the bubble are heated is made.
    The equations of evolution of the bubble radius thus obtained will be considered as the equations for evolution of the
    `scale factor' of the Universe.\\

    It is hoped that the dynamics obtained helps to shed light on modeling of inflation, dark matter/energy etc.

    We also attempt to find the equation of state of the matter in the Universe if it is assumed to evolve according
    to the dynamics obtained. These equations of state are analysed to find hints of dark matter, dark energy and the like.\\[20pt]

    \subsection{General heating of the bubble}

    Consider a spherical bubble containing an ideal gas in a surrounding medium of some liquid - e.g. water - at a

    pressure and temperature $P_0$ and $T_0$ . Let the surface tension at the interface be S and for the gas in the bubble
    let $\frac{C_P}{R_N} = \alpha$. At an instant of time say $t = 0$ heat from a power source -- like a laser beam or a spark plug --
     is incident inside the bubble causing it to expand. Let us consider the dynamics of such a bubble-\\

    Now for any radius R of the bubble we have for the gas inside.

        \begin{equation}\label{dP1}
        P = P_0 + \frac{2S}{R}\ \Rightarrow\ \ dP = - \frac{2S}{R^2}dR
        \end{equation}\\[10pt]

    From the conservation of energy for the bubble we have--

        \begin{equation}\label{dQ1}
        dQ = nC_vdT + SdA + PdV
        \end{equation}\\[10pt]

    We also have for the bubble--\\[10pt]

        \[
         A = 4\pi R^2\ \Rightarrow\ \ dA = 8\pi RdR
        \]
        \begin{equation}\label{dAdV}
        \end{equation}
        \[
         V = \frac{4}{3}\pi R^3\ \Rightarrow\ \ dV = 4\pi R^2dR
        \]\\[10pt]

    For the ideal gas in the bubble we have--\\[10pt]

        \begin{equation}\label{dT}
        T = \frac{PV}{nR_N}\ \Rightarrow\ \ dT = \frac{PdV + VdP}{nR_N}
        \end{equation}\\[20pt]

    Using eqn (\ref{dT}) in (\ref{dQ1}) --\\[10pt]

        \[
         dQ = \frac{C_V}{R_N}(PdV + VdP) + SdA + PdV
        \]
        \begin{equation}\label{dQ2}
        \     \ = (PdV)\alpha + (VdP)(\alpha - 1) + SdA
        \end{equation}\\[10pt]

    Now we substitute eqns (\ref{dP1}) and (\ref{dAdV}) in (\ref{dQ2})\\[10pt]

        \[
         dQ = \alpha (P_0 + \frac{2S}{R})(4\pi R^2dR) + (\alpha - 1)(\frac{4}{3}\pi R^3)(- \frac{2S}{R^2}dR) + S(8\pi RdR)
        \]\\[10pt]

     Dividing throughout by dt and using $W(t) = \frac{dQ}{dt}$  --\\[10pt]

        \[
         W(t) = \dot{R} \left(\alpha (P_0 + \frac{2S}{R})(4\pi R^2) + (\alpha - 1)(\frac{4}{3}\pi R^3)(- \frac{2S}{R^2}) + S(8\pi R) \right)
        \]

    i.e -- \\[10pt]

        \begin{equation}\label{R.1}
        \dot{R} = \frac{W(t)}{4\pi R\left( \alpha P_0R + \frac{4S}{3}(\alpha+2)\right)}
        \end{equation}

\newpage

    \subsection{Case1 - Constant power source}

    We consider a special case of eqn (\ref{R.1}) when the power supplied is a constant.
    So we have --\\
        \begin{equation}\label{Wconst}
        W(t) = W_0
        \end{equation}\\

    Then we see that we get the following equation which can be readily integrated--\\
        \[
         \int 4\pi R\left( \alpha P_0R + \frac{4S}{3}(\alpha+2)\right)dR = W_0\int dt
        \]\\
    to give--\\
        \begin{equation}\label{R-teqn1}
        \frac{R^2}{3}\left(\alpha P_0R + 2S(\alpha+2)\right) = \frac{W_0}{4\pi}t + \kappa
        \end{equation}\\[10pt]

\newpage
    \subsection{Case2 - Bubble conducting heat}

    Now lets consider a special case, where the bubble conducts heat across its surface to the surrounding reservoir.
    Assume that the reservoir is large enough so that it is unaffected by the heat absorbed or released by the bubble.\\

    Let the rate of heat conduction be proportional to the area and the temperature difference across the bubble surface.
    So we have --\\[10pt]

        \begin{equation}\label{Wcooling}
        W = W_C = - \beta R^2(T - T_0)
        \end{equation}\\[10pt]

    We substitute for the T's using the eqns (\ref{dP1}) (\ref{dAdV}) and (\ref{dT}) -- \\[10pt]

        \[
         T = \frac{1}{nR_N}(P_0 + \frac{2S}{R})(\frac{4}{3}\pi R^3)
        \]
        \begin{equation}\label{Tcooling}
        \   \ = \frac{4\pi}{3nR_N}(P_0R^3 + 2SR^2)
        \end{equation}\\[10pt]

    Using eqns (\ref{Wcooling}) and (\ref{Tcooling}) we get-- \\[10pt]

        \[
         W_C = - \frac{4\pi}{3nR_N}\beta R^2 \left( P_0R^3 + 2SR^2 - P_0R_0^3 - 2SR_0^2\right)
        \]

    which reduces to --

        \[
         W_C = - \frac{4\pi}{3nR_N}\beta R^2 \left(P_0(R^3 - R_0^3) + 2S(R^2 - R_0^2)\right)
        \]

        $\Rightarrow$

        \begin{equation}\label{Wcooling1}
        W_C = - \frac{4\pi}{3nR_N}\beta R^2 (R - R_0) \left(P_0R^2 + (P_0R_0 + 2S)(R + R_0)\right)
        \end{equation}\\[10pt]

    Using (\ref{Wcooling1}) in (\ref{R.1})\\[10pt]

        \begin{equation}\label{R.cooling}
        \dot{R} = -\frac{\beta}{3nR_N}R(R-R_0)\frac{P_0R^2 + (P_0R_0 + 2S)(R + R_0)}{(\alpha P_0)R + \frac{4S}{3}(\alpha + 2)}
        \end{equation}\\
We note that the sudden increase in $R(t)$  is \emph{very similar to
the inflation scenario in conventional Cosmology.}

\newpage

    \subsection{Other models}

     We shall consider variants of previously developed models and compare them to existing cosmological models.\\

    \subsubsection{Hybrid model with inflation\\[10pt]}
     We consider a model where the bubble first expands rapidly according to eqn (\ref{R.cooling}) but then eqn (\ref{R-teqn1}) takes over.
     This involves a discontinuous change in the function $W(t)$ from that given by eqn (\ref{Wcooling}) to eqn (\ref{Wconst}).
     This can be considered to be \emph{analogous to a phase transition that is presumed to have happened during inflation}.
     \\[10pt]

\subsubsection{Single model with inflation\\[10pt]}
     We shall use a variant of eqn (\ref{R.cooling}) where $R_0$ is no longer considered a constant but a function of time.
     We note that in such a case $R(t)$ initially undergoes a sudden \emph{inflation} and then follows exactly the functional
     form given by $R_0(t)$.\\

    Thus we use the Friedmann evolution as the form for $R_0(t)$ and so we have an \emph{initial inflation followed by a Friedmann evolution!!!}\\

    \section{Gravitational Waves\\[15pt]}

    \subsection*{Introduction}
    Here we shall attempt to solve the equations of General Relativity for the case of small perturbations in the background metric
    and obtain wave solutions.\\

    Calculations are done upto $1^{st}$ and $2^{nd}$ order in the perturbations.\\

    We hypothesize that \emph{a primordial gravitational wave is responsible for the anisotropy in the CMBR as observed by WMAP}.
     Starting from a isotropic primordial gravitational wave, we hope to make use of the non-linear coupling in a gravitational
      wave solution --- \emph{upto $2^{nd}$ order} --- to evolve the wave into the anisotropies seen in the WMAP data.\\[10pt]

    \subsection{Equations of General Relativity}

    We list here the equations of General Relativity that shall be useful in this calculation.\\

    Let us consider a metric $g_{ik}$ on spacetime; then we have the following --\\[10pt]

    \begin{itemize}
        \item   The Christoffel symbols are\\
            \begin{equation}\label{christ.def}
            {\Gamma ^i}_{kl} = \frac{1}{2}g^{im}(g_{mk,l} + g_{ml,k} - g_{lk,m})
            \end{equation}\\

        \item   The Riemann tensor is\\
            \begin{equation}\label{reimann.def}
            {R^i}_{klm} = {\Gamma ^i}_{km,l} - {\Gamma ^i}_{kl,m} + {\Gamma ^i}_{nl}{\Gamma ^n}_{km} - {\Gamma ^i}_{nm}{\Gamma ^n}_{kl}
            \end{equation}\\

        \item   The Riemann tensor has following symmetries\\
            \[
             R_{iklm} = - R_{kilm} = - R_{ikml} = R_{lmik}
            \]\\
            \begin{equation}\label{reimann.sym}
            R_{iklm} + R_{ilmk} + R_{imkl} = 0
            \end{equation}\\
            \[
             {R^n}_{ikl;m} + {R^n}_{ilm;k} + {R^n}_{imk;l} = 0
            \]\\

        \item   The Ricci tensor is given by\\
            \begin{equation}\label{ricci.def}
            R_{ik} = {R^l}_{ilk}
            \end{equation}\\

        \item   The scalar curvature is\\
            \begin{equation}\label{scalecurv.def}
            R = {R^i}_i = g^{ik}R_{ik}
            \end{equation}\\

        \item   Given the energy-momentum tensor $T_{ik}$ Einstein equation can be written  as\\
            \[
             R_{ik} - \frac{1}{2}g_{ik}R = \frac{8\pi k}{c^4}T_{ik}
            \]\\
            \begin{equation}\label{einstein.def}
            {R_i}^k - \frac{1}{2}{\delta_i}^kR = \frac{8\pi k}{c^4}{T_i}^k
            \end{equation}\\
            \[
             R_{ik} = \frac{8\pi k}{c^4}(T_{ik} - \frac{1}{2}g_{ik}T)
            \]\\

    \end{itemize}

\newpage

    \subsection{Perturbations}

    To analyze the phenomena of gravitational waves, a perturbation \(h_{ik}\) in the metric of spacetime \(g_{ik}\) over its
     ``background'' value \(g_{ik}^{(0)}\) is considered as --
        \begin{equation}\label{met_perturbation}
        g_{ik} = {g_{ik}}^{(0)}+h_{ik}
        \end{equation}\\

    It is chosen by convention to ``raise'' and ``lower'' indices using the ``background'' metric \(g_{ik}^{(0)}\).
    That is - \(h_k^m = g^{mi(0)}h_{ik}\), where --
        \begin{equation}
        {g^{mi}}^{(0)}{g_{ik}}^{(0)}=\delta_k^m
        \end{equation}\\

    The perturbations in the contravariant metric is evaluated by solving the equation --
        \begin{equation}
        g^{mi}g_{ik}=\delta_k^m
        \end{equation}\\

    Let -- \(g^{mi} = {g^{mi}}^{(0)}+f^{mi}\). Then substitution into above equation gives --
        \[
         \left({g^{mi}}^{(0)}+f^{mi}\right)\left({g_{ik}}^{(0)}+h_{ik}\right) = \delta_k^m
        \]

    i.e.
        \begin{equation}\label{pert_eqn}
         h_k^m + f_k^m + f^{mi}h_{ik}=0
        \end{equation}

    Now, upto first order in \(h\) we have --

        \[
         {f^{mk}}^{(1)}=-h^{mk}
        \]\\

    Substituting this back in Eq.(\ref{pert_eqn}) we find the next order as --

        \[
         {f^{mk}}^{(2)}= h^{mi}h_i^k
        \]\\

    Similarly we can iteratively find the higher orders of \(f\) as --

        \begin{equation}
        {f^{mk}}^{(n)} = -{f^{mi}}^{(n-1)}h_i^k \hspace{10pt};\hspace{10pt}{f^{mk}}^{(1)}=-h^{mk}
        \end{equation}

\subsection{Ricci Tensor}

    The Riemann tensor is given as --
        \begin{equation}\label{riemann_defn}
        {R^i}_{klm} = \frac{1}{2}g^{in}\left(g_{nm,kl}+g_{kl,nm}-g_{nl,km}-g_{km,nl}\right)
        \end{equation}

    The Ricci tensor is thus --
        \begin{equation}
        R_{km}={R^l}_{klm} = \frac{1}{2}g^{ln}\left(g_{nm,kl}+g_{kl,nm}-g_{nl,km}-g_{km,nl}\right)
        \end{equation}\\

    The Ricci tensor is now computed upto \(2^{nd}\) order in the perturbation \(h\).

        \[
         \frac{1}{2}\left({g^{ln}}^{(0)}-h^{ln}+h^{lp}h_p^n\right)
        \left(
            \begin{array}{c}
            {g_{nm,kl}}^{(0)}+{g_{kl,nm}}^{(0)}-{g_{nl,km}}^{(0)}-{g_{km,nl}}^{(0)} \\
            + h_{nm,kl}+h_{kl,nm}-h_{nl,km}-h_{km,nl}
            \end{array}
        \right)
        \]\\

    Now separating the Ricci tensor into parts of different orders we have --

        \begin{equation}\label{ricci0_defn}
         {R_{km}}^{(0)} = \frac{1}{2}{g^{ln}}^{(0)}\left({g_{nm,kl}}^{(0)}+{g_{kl,nm}}^{(0)}-{g_{nl,km}}^{(0)}-{g_{km,nl}}^{(0)}\right)
        \end{equation}\\

        \[
        {R_{km}}^{(1)} = \frac{1}{2}{g^{ln}}^{(0)}\left(h_{nm,kl}+h_{kl,nm}-h_{nl,km}-h_{km,nl}\right)
        \]
        \begin{equation}\label{ricci1_defn}
        \hspace{70pt}- \frac{1}{2}h^{ln}\left({g_{nm,kl}}^{(0)}+{g_{kl,nm}}^{(0)}-{g_{nl,km}}^{(0)}-{g_{km,nl}}^{(0)}\right)
        \end{equation}\\

        \[
         {R_{km}}^{(2)} = - \frac{1}{2}h^{ln}\left(h_{nm,kl}+h_{kl,nm}-h_{nl,km}-h_{km,nl}\right)
        \]
        \begin{equation}\label{ricci2_defn}
        \hspace{100pt}+\frac{1}{2}h^{lp}h_p^n\left({g_{nm,kl}}^{(0)}+{g_{kl,nm}}^{(0)}-{g_{nl,km}}^{(0)}-{g_{km,nl}}^{(0)}\right)
        \end{equation}

\newpage

    \subsection{$1^{st}$ order calculation with a Minkowskian background\\[10pt]}

    Now we shall now proceed with the calculations with following assumptions -- \\
        \begin{enumerate}
        \item   The unperturbed background spacetime is Minkowskian.\\[8pt]
            \( \Rightarrow\ {g_{ik}}^{(0)} =  \eta _{ik} \equiv
            \) Minkowskian metric\\
        \item   Only terms upto $1^{st}$ order in $h_{ik}$ or its derivatives are       significant.
        All    higher order terms are negligible.
        \end{enumerate}\ \ \\[10pt]

    For a Minkowskian background we also have ${{R^i}_{klm}}^{(0)} = 0$; so we need to consider only the $1^{st}$
    order term of the Riemann tensor\\.
    From Eq. (\ref{ricci1_defn}), the Ricci tensor of \(1^{st}\) order is --\\

            \begin{equation}\label{ricci_minkow}
            {R_{km}}^{(1)} = \frac{1}{2}\eta^{ln}\left(h_{nm,kl}+h_{kl,nm}-h_{nl,km}-h_{km,nl}\right)
            \end{equation}

    which reduces to --

        \begin{equation}
        R_{km} = \frac{1}{2} \Box h_{km} + \frac{1}{2}(h^l_{m,k,l} - h_{,k,m} + {{h_{kl}}^{,l}}_{,m})
        \end{equation}\\

    where $\Box$ denotes the d'Alembertian operator:
        \[
         \Box = -\eta ^{lm}\frac{\partial ^2}{\partial x^l\partial x^m} = \bigtriangleup - \frac{\partial ^2}{c^2\partial t^2}
        \]\\[20pt]

    To simplify the equation, we make use of the gauge freedom as observed in Eq.s (\ref{transform.x}) and (\ref{transform.h})\\.

    We impose on $h_{ik}$ the following additional condition--

        \[
         \psi _i^k = h_i^k - \frac{1}{2} \delta _i^k h
        \]
        \begin{equation}\label{gaugecond}
        \psi ^k_{i,k} = 0
        \end{equation}\\

    This is known as the \emph{harmonic gauge} or the \emph{de Donder
    gauge}.

    We again note that the above condtions still \emph{do not} imply a preferred frame of reference.
     Substituting from Eq. (\ref{transform.h}) we can see that condition (\ref{gaugecond}) allows for
     co-ordinate transformations as in Eq. (\ref{transform.x}) provided--
        \begin{equation}
        \Box \xi^i = 0
        \end{equation}\\[10pt]

    The Eq. (\ref{gaugecond}) gives us--
        \[
         h^k_{i,k} = \frac{1}{2}\delta ^k_ih_{,k} = \frac{1}{2}h_{,i}
        \]\\

    Using this we get--
        \[
         h^l_{m,k,l} - h_{,k,m} + {{h_{kl}}^{,l}}_{,m} = 0
        \]\\

    And Eq. (\ref{ricci.1.h}) reduces to --
        \begin{equation}\label{ricci.1.box}
        R_{km} = \frac{1}{2} \Box h_{km}
        \end{equation}\\[15pt]

    For the case of vacuum we have
        \[
         R_{km} = 0
        \]\\
    and so Eq. (\ref{ricci.1.box}) gives us the usual wave equation--
        \begin{equation}\label{waveEq..1}
        \Box h_{km} = 0
        \end{equation}\ \\[20pt]

\newpage

    \subsection{$2^{nd}$ order calculation}

    The wave equation obtained in the previous section is \emph{linear}. As a result there is no self-interaction
    in the wave and so \emph{one cannot hope to evolve an isotropic wave into an anisotropic one}.
    With this purpose in mind we here attempt to calculate the $2^{nd}$ order corrections to the above equations.\\[15pt]

    The second order Ricci tensor is given by Eq. (\ref{ricci2_defn})--\\
            \begin{equation}
            {R_{km}}^{(2)} = -\frac{1}{2}h^{ln}\left(h_{nm,kl}+h_{kl,nm}-h_{nl,km}-h_{km,nl}\right)
            \end{equation}

    Thus the wave equation in vacuum reduces to --

            \begin{equation}
            \Box h_{km} = h^{ln}\left(h_{nm,kl}+h_{kl,nm}-h_{nl,km}-h_{km,nl}\right)
            \end{equation}

\section{Perturbations}

    To analyse the phenomena of gravitational waves, a perturbation \(h_{ik}\) in the metric of spacetime \(g_{ik}\) over
     its ``background'' value \(g_{ik}^{(0)}\) is considered as --
        \begin{equation}\label{met_perturbation}
        g_{ik} = {g_{ik}}^{(0)}+h_{ik}
        \end{equation}\\

    It is chosen by convention to ``raise'' and ``lower'' indices using the ``background'' metric \(g_{ik}^{(0)}\).
    That is - \(h_k^m = g^{mi(0)}h_{ik}\), where --
        \begin{equation}
        {g^{mi}}^{(0)}{g_{ik}}^{(0)}=\delta_k^m
        \end{equation}\\

    The perturbations in the contravariant metric is evaluated by solving the equation --
        \begin{equation}
        g^{mi}g_{ik}=\delta_k^m
        \end{equation}\\

    Let -- \(g^{mi} = {g^{mi}}^{(0)}+f^{mi}\). Then substitution into above equation gives --
        \[
         \left({g^{mi}}^{(0)}+f^{mi}\right)\left({g_{ik}}^{(0)}+h_{ik}\right) = \delta_k^m
        \]

    i.e.
        \begin{equation}\label{pert_eqn}
         h_k^m + f_k^m + f^{mi}h_{ik}=0
        \end{equation}

    Now, upto first order in \(h\) we have --

        \[
         {f^{mk}}^{(1)}=-h^{mk}
        \]\\

    Substituting this back in Equation(\ref{pert_eqn}) we find the next order as --

        \[
         {f^{mk}}^{(2)}= h^{mi}h_i^k
        \]\\

    Similarly we can iteratively find the higher orders of \(f\) as --

        \begin{equation}
        {f^{mk}}^{(n)} = -{f^{mi}}^{(n-1)}h_i^k \hspace{10pt};\hspace{10pt}{f^{mk}}^{(1)}=-h^{mk}
        \end{equation}

\newpage

    \section{Christoffel Symbols}

    The Christoffel symbols in terms of the metric tensor are --
        \begin{equation}\label{christ_defn}
        \Gamma^n_{km} = \frac{1}{2}g^{in}\left(g_{mi,k}+g_{ki,m}-g_{km,i}\right)
        \end{equation}

    where the `comma' (,) denotes partial derivatives.\\[10pt]

    Now, upto \(2^{nd}\) order in the metric perturbations --

        \[
         \Gamma^n_{km}= \frac{1}{2}\left({g^{in}}^{(0)}-h^{in}+h^i_ph^{pn}\right)
        \left(
            \begin{array}{c}
            {g_{mi,k}}^{(0)}+{g_{ki,m}}^{(0)}-{g_{km,i}}^{(0)} \\
            h_{mi,k}+h_{ki,m}-h_{km,i}
            \end{array}
        \right)
        \]\\[15pt]

    So the Christoffel symbols of different orders are --

        \begin{equation}\label{christ0_defn}
        {\Gamma^n_{km}}^{(0)} = \frac{1}{2}{g^{in}}^{(0)}\left({g_{im,k}}^{(0)}+{g_{ik,m}}^{(0)}-{g_{km,i}}^{(0)}\right)
        \end{equation}

        \[
         {\Gamma^n_{km}}^{(1)} = -\frac{1}{2}h^{in}\left({g_{im,k}}^{(0)}+{g_{ik,m}}^{(0)}-{g_{km,i}}^{(0)}\right)
        \]
        \begin{equation}\label{christ1_defn}
        \hspace{20pt}+ \frac{1}{2}{g^{in}}^{(0)}\left(h_{im,k}+h_{ik,m}-h_{km,i}\right)
        \end{equation}

        \[
         {\Gamma^n_{km}}^{(2)} = \frac{1}{2}h^i_ph^{pn}\left({g_{im,k}}^{(0)}+{g_{ik,m}}^{(0)}-{g_{km,i}}^{(0)}\right)
        \]
        \begin{equation}\label{christ2_defn}
        \hspace{90pt}-\frac{1}{2}h^{in}\left(h_{im,k}+h_{ik,m}-h_{km,i}\right)
        \end{equation}

\newpage
        \section{Harmonic Gauge}

        To find specific solutions of the Einstein equation, the freedom in its general covariance is restricted by
        imposing certain additional conditions on the metric called \emph{gauge conditions}.

        The \emph{harmonic gauge}, generally used in analysis of gravitational waves, is specified by the conditions --
            \begin{equation}\label{harmonicgauge_defn}
            H^n = g^{km}\Gamma^n_{km} = 0
            \end{equation}\\

        We now write these conditions for different orders of \(h\) -- \\

            \[
            {H^n}^{(0)} = {g^{km}}^{(0)}{\Gamma^n_{km}}^{(0)}
            \]

            \begin{equation}\label{harmonic0}
            =\frac{1}{2}{g^{km}}^{(0)}{g^{in}}^{(0)}\left({g_{im,k}}^{(0)}+{g_{ik,m}}^{(0)}-{g_{km,i}}^{(0)}\right)
            \end{equation}\\[15pt]

            \[
             {H^n}^{(1)} = {g^{km}}^{(0)}{\Gamma^n_{km}}^{(1)} + {g^{km}}^{(1)}{\Gamma^n_{km}}^{(0)}
            \]

            \[
             = -\frac{1}{2}\left({g^{in}}^{(0)}h^{km}+{g^{km}}^{(0)}h^{in}\right)\left({g_{im,k}}^{(0)}+{g_{ik,m}}^{(0)}-{g_{km,i}}^{(0)}\right)
            \]

            \begin{equation}\label{harmonic1}
            + \frac{1}{2}{g^{km}}^{(0)}{g^{in}}^{(0)}\left(h_{im,k}+h_{ik,m}-h_{km,i}\right)
            \end{equation}\\[15pt]

            \[
             {H^n}^{(2)} = {g^{km}}^{(0)}{\Gamma^n_{km}}^{(2)} + {g^{km}}^{(2)}{\Gamma^n_{km}}^{(0)} + {g^{km}}^{(1)}{\Gamma^n_{km}}^{(1)}
            \]

            \[
             = \frac{1}{2}\left({g^{km}}^{(0)}h^i_ph^{pn}+{g^{in}}^{(0)}h^k_ph^{pm}+h^{km}h^{in}\right)
             \left({g_{im,k}}^{(0)}+{g_{ik,m}}^{(0)}-{g_{km,i}}^{(0)}\right)
            \]

            \begin{equation}\label{harmonic2}
            -\frac{1}{2}\left({g^{in}}^{(0)}h^{km}+{g^{km}}^{(0)}h^{in}\right)\left(h_{im,k}+h_{ik,m}-h_{km,i}\right)
            \end{equation}

\newpage

    \section{Ricci Tensor}

    The Riemann tensor is given as --\\

        \begin{equation}\label{riemann_defn}
        R_{iklm} = \frac{1}{2}\left(g_{kl,mi}+g_{mi,kl}-g_{il,km}-g_{km,il}\right)  + g_{jp}
        \left(\Gamma_{kl}^j\Gamma_{mi}^p-\Gamma_{km}^j\Gamma_{il}^p\right)
        \end{equation}\\[10pt]

    The Ricci tensor is thus --\\
        \begin{equation}\label{riccidefn}
        R_{km}={R^l}_{klm} = \frac{1}{2}g^{ln}\left(g_{mn,kl}+g_{kl,mn}-g_{nl,km}-g_{km,nl}\right) + g^{ln}g_{jp}
        \left(\Gamma_{kl}^j\Gamma_{mn}^p-\Gamma_{km}^j\Gamma_{nl}^p\right)
        \end{equation}\\[20pt]

    Now in the case of a Minkowskian background, we have \({g_{ik}}^{(0)} = \eta_{ik}\). So \({R_{km}}^{(0)} = 0\)
    and for first order we have --

        \[
         {R_{km}}^{(1)} = \frac{1}{2}\eta^{ln}\left(h_{kl,mn}+h_{mn,kl}-h_{nl,km}-h_{km,nl}\right)
        \]

        \begin{equation}
        = \frac{1}{2}\left(h^n_{k,nm}+h^l_{m,lk}-h_{,km}+\Box h_{km}\right)
        \end{equation}\\

        Also from Equation (\ref{harmonic1}) the harmonic gauge condition becomes --

        \begin{equation}\label{harmonic1_1}
        h^l_{k,l} - \frac{1}{2}h_{,k} = 0
        \end{equation}\\

    and thus we have --

        \begin{equation}
        {R_{km}}^{(1)} = \frac{1}{2}\Box h_{km}
        \end{equation}

    which in the first order approximation reduces to the ordinary wave equation.\\[20pt]

    Also the second order Ricci tensor is --

        \[
         -\frac{1}{2}h^{ln}\left(h_{kl,mn}+h_{mn,kl}-h_{km,nl}-h_{nl,km}\right)
        \]

        \begin{equation}
        + \frac{1}{4}\eta^{ln}\eta_{jp}
        \left[
            \begin{array}{c}
            \left(h^j_{k,l}+h^j_{l,k}-{h_{kl}}^{,j}\right)\left(h^p_{m,n}+h^p_{n,m}-{h_{mn}}^{,p}\right)\\
            -\left(h^j_{k,m}+h^j_{m,k}-{h_{km}}^{,j}\right)\left(h^p_{l,n}+h^p_{n,l}-{h_{nl}}^{,p}\right)
            \end{array}
        \right]
        \end{equation}\\

    Using the harmonic gauge as in Equation.(\ref{harmonic1_1}) we get --

        \[
         \eta^{ln}\left(h^p_{l,n}+h^p_{n,l}-{h_{nl}}^{,p}\right) = 0
        \]

    and the second term reduces to --

        \[
         \frac{1}{4}\left({h_{pk}}^{,n}+h^n_{p,k}-h^n_{k,p}\right)\left(h^p_{m,n}+h^p_{n,m}-{h_{mn}}^{,p}\right)
        \]

    Now we use the condition of Ricci-flatness \(R_{km} = 0\) and thus we get the nonlinear wave equation --

        \[
         \Box h_{km } = h^{ln}\left(h_{kl,mn}+h_{mn,kl}-h_{km,nl}-h_{nl,km}\right)
        \]

        \begin{equation}\label{wave_nonlinear}
        \hspace{80pt}-\frac{1}{2}\left({h_{pk}}^{,n}+h^n_{p,k}-h^n_{k,p}\right)\left(h^p_{m,n}+h^p_{n,m}-{h_{mn}}^{,p}\right)
        \end{equation}\\

\newpage

    \section{The Wave Equations}

    Now lets us assume that the metric perturbation \(h\) takes a form similar to the linear solution of a transverse
     wave traveling along the Z - axis. So we can write \(h_{ik}\) as --

        \begin{equation}
        h_{ik} = \left(
                \begin{array}{cccc}
                k & 0 & 0 & 0 \\
                0 & a & b & 0 \\
                0 & b & -a& 0 \\
                0 & 0 & 0 & 0 \\
                \end{array}
            \right)
        \end{equation}

    where \(a ,b, k\) are function of \(x^k\).\\

    For the case of Minkowski background, the zeroth order harmonic condition is satisfied identically.
     We thus, evaluate the harmonic gauge conditions of \(1^{st}\) order --\\

        \[
         H^{n(1)} = h^{0n}_{,0}+h^{1n}_{,1}+h^{2n}_{,2}-\frac{1}{2}\left(h^{0,n}_0+h^{1,n}_1+h^{2,n}_2\right)
        \]
        \[
         = h^{0n}_{,0}+h^{1n}_{,1}+h^{2n}_{,2}-\frac{1}{2}h^{0,n}_0
        \]\\

    Thus for different values \(n=0,1,2,3\), we get --

        \begin{equation}\label{harmonic1_minkowski_0}
        {H^0}^{(1)}= \frac{1}{2}\frac{\partial k}{\partial t} = 0
        \end{equation}

        \begin{equation}\label{harmonic1_minkowski_1}
        {H^1}^{(1)}= \frac{\partial a}{\partial x}+\frac{\partial b}{\partial y}+\frac{1}{2}\frac{\partial k}{\partial x} = 0
        \end{equation}

        \begin{equation}\label{harmonic1_minkowski_2}
        {H^2}^{(1)}= \frac{\partial b}{\partial x}-\frac{\partial a}{\partial y}+\frac{1}{2}\frac{\partial k}{\partial y} = 0
        \end{equation}

        \begin{equation}\label{harmonic1_minkowski_3}
        {H^3}^{(1)}= \frac{1}{2}\frac{\partial k}{\partial z} = 0
        \end{equation}

    These equations will help simplify the wave equation further.\\[15pt]

    The wave equation can now be written for each component of \(h_{ik}\). For \(\mu = 1,2,3\), we have \(h_{0\mu}=0\),
    and using the first and last harmonic conditions we have --

        \[
         0 = h^{11}\left(h_{\mu 1,01}-h_{11,0\mu}\right)+h^{12}\left(h_{\mu 2,01}-h_{12,0\mu}\right)+h^{21}
         \left(h_{\mu 1,02}-h_{21,0\mu}\right)+h^{22}\left(h_{\mu 2,02}-h_{22,0\mu}\right)
        \]
        \[
         -\frac{1}{2}\left(
                    \begin{array}{c}
                    h_{00}^{,1}\left(-{h_{\mu 1}}^{,0}\right)+h_{00}^{,2}\left(-{h_{\mu 2}}^{,0}\right)-h^0_{0,1}
                    \left(h^1_{\mu ,0}\right)-h^0_{0,2}\left(h^2_{\mu ,0}\right)\\[10pt]
                    +h^1_{1,0}\left(h^1_{\mu ,1}+h^1_{1,\mu}-{h_{\mu 1}}^{,1}\right)+h^1_{2,0}
                    \left(h^2_{\mu ,1}+h^2_{1,\mu}-{h_{\mu 1}}^{,2}\right)\\[10pt]
                    +h^2_{1,0}\left(h^1_{\mu ,2}+h^1_{2,\mu}-{h_{\mu 2}}^{,1}\right)+h^2_{2,0}
                    \left(h^2_{\mu ,2}+h^2_{2,\mu}-{h_{\mu 2}}^{,2}\right)
                    \end{array}
                  \right)
        \]\\[20pt]

    Now for \(\mu = 1\), we get --

        \[
        b\frac{\partial}{\partial t}\left(\frac{\partial a}{\partial y}-\frac{\partial b}{\partial x}\right) - a\frac{\partial}{\partial t}
        \left(\frac{\partial a}{\partial x}+\frac{\partial b}{\partial y}\right)
        \]
        \[
        -\left(\frac{\partial k}{\partial x}\frac{\partial a}{\partial t}+\frac{\partial k}{\partial y}\frac{\partial b}{\partial t}\right)-
        \left(\frac{\partial a}{\partial x}\frac{\partial a}{\partial t}+\frac{\partial b}{\partial x}\frac{\partial b}{\partial t}\right) = 0
        \]\\

    On simplification using the harmonic conditions, we have --

        \begin{equation}\label{23eqn_1}
        \frac{\partial a}{\partial t}\left(2\frac{\partial b}{\partial y}+\frac{\partial a}{\partial x}\right)=
        \frac{\partial b}{\partial t}\left(2\frac{\partial a}{\partial y}-\frac{\partial b}{\partial x}\right)
        \end{equation}\\[15pt]

    Similarly for \(\mu =2\) we have--

        \begin{equation}\label{23eqn_2}
        \frac{\partial b}{\partial t}\left(2\frac{\partial a}{\partial x}+\frac{\partial b}{\partial y}\right)=
        \frac{\partial a}{\partial t}\left(2\frac{\partial b}{\partial x}-\frac{\partial a}{\partial y}\right)
        \end{equation}\\[15pt]

    and for \(\mu = 3\) --

        \begin{equation}\label{23eqn_3}
        a\frac{\partial ^2a}{\partial t\partial z}+b\frac{\partial ^2b}{\partial t\partial z} =
        -\frac{1}{2}\left(\frac{\partial a}{\partial t}\frac{\partial a}{\partial z}+\frac{\partial b}{\partial t}\frac{\partial b}{\partial z}\right)
        \end{equation}\\[20pt]

    For \(\mu = 1,2,3\), \(h_{3\mu}=0\) so we have --

        \[
         0 = h^{11}\left(h_{\mu1,31}-h_{11,3\mu}\right)+h^{12}\left(h_{\mu2,31}-h_{21,3\mu}\right)+h^{21}\left(h_{\mu1,32}-h_{12,3\mu}\right)
         +h^{22}\left(h_{\mu2,32}-h_{22,3\mu}\right)
        \]
        \[
         -\frac{1}{2}\left(
                    \begin{array}{c}
                    h^1_{1,3}\left(h^1_{\mu,1}+h^1_{1,\mu}-{h_{\mu1}^{,1}}\right)+h^1_{2,3}\left(h^2_{\mu,1}+h^2_{1,\mu}-{h_{\mu1}^{,2}}\right)\\[10pt]
                    h^2_{1,3}\left(h^1_{\mu,2}+h^1_{2,\mu}-{h_{\mu2}^{,1}}\right)+h^2_{2,3}\left(h^2_{\mu,2}+h^2_{2,\mu}-{h_{\mu2}^{,2}}\right)\\[10pt]
                    \end{array}
                  \right)
        \]\\[20pt]

    Now for \(\mu = 1\)

        \[
        b\frac{\partial}{\partial z}\left(\frac{\partial a}{\partial y}-\frac{\partial b}{\partial x}\right) - a\frac{\partial}{\partial z}
        \left(\frac{\partial a}{\partial x}+\frac{\partial b}{\partial y}\right)
        -\frac{\partial a}{\partial z}\frac{\partial a}{\partial x}-\frac{\partial b}{\partial z}\frac{\partial b}{\partial x}=0
        \]\\

    Using the harmonic conditions we have --

        \begin{equation}\label{23eqn_4}
        \frac{\partial a}{\partial z}\frac{\partial a}{\partial x}=-\frac{\partial b}{\partial z}\frac{\partial b}{\partial x}
        \end{equation}

    Similarly for \(\mu = 2\)

        \begin{equation}\label{23eqn_5}
        \frac{\partial a}{\partial z}\frac{\partial a}{\partial y}=-\frac{\partial b}{\partial z}\frac{\partial b}{\partial y}
        \end{equation}

    and \(\mu =3\)

        \begin{equation}\label{23eqn_6}
        a\frac{\partial^2 a}{\partial z^2}+b\frac{\partial^2 b}{\partial z^2} = -\frac{1}{2}\left[\left(\frac{\partial a}{\partial z}\right)^2+
        \left(\frac{\partial b}{\partial z}\right)^2\right]
        \end{equation}\\[20pt]

    Similarly writing the components explicitly we get for \(h_{11} = a\)  --

        \[
         \Box a  = -k\left(\frac{\partial^2k}{\partial x^2}+\frac{\partial^2a}{\partial t^2}\right) -
         a \left(2 \frac{\partial^2b}{\partial x\partial y}+\frac{\partial^2a}{\partial x^2}-\frac{\partial^2a}{\partial y^2}\right)
        \]
        \[
         -\left[\left(\frac{\partial a}{\partial x}\right)^2+\left(\frac{\partial a}{\partial y}\right)^2+
         \left(\frac{\partial a}{\partial z}\right)^2\right] + \left(\frac{\partial b}{\partial x}\right)
         \left(\frac{\partial k}{\partial y}\right)-\left(\frac{\partial b}{\partial z}\right)^2
        \]
        \begin{equation}
        +\left[\left(\frac{\partial a}{\partial t}\right)^2+\left(\frac{\partial b}{\partial t}\right)^2\right]-\frac{1}{2}
        \left(\frac{\partial k}{\partial x}\right)^2
        \end{equation}\\[20pt]

    For \(h_{12} = b\)  --

        \[
         \Box b = -k\left(\frac{\partial^2k}{\partial x\partial y}+\frac{\partial^2b}{\partial t^2}\right) -
         b \left(2 \frac{\partial^2b}{\partial x\partial y}+\frac{\partial^2a}{\partial x^2}-\frac{\partial^2a}{\partial y^2}\right)
        \]
        \begin{equation}
         -2\left(\frac{\partial a}{\partial x}\right)\left(\frac{\partial a}{\partial y}\right) +
         \left(\frac{\partial a}{\partial x}\right)\left(\frac{\partial b}{\partial x}\right) -
         \left(\frac{\partial a}{\partial y}\right)\left(\frac{\partial b}{\partial y}\right) -
         \frac{1}{2}\left(\frac{\partial k}{\partial x}\right)\left(\frac{\partial k}{\partial y}\right)
        \end{equation}\\[20pt]

    For \(h_{21} = b\) we get the same equation as above as is to be expected.

    For \(h_{22} = -a\)  --

        \[
         \Box a  = -k\left(\frac{\partial^2k}{\partial y^2}+\frac{\partial^2a}{\partial t^2}\right) -
         a \left(2 \frac{\partial^2b}{\partial x\partial y}+\frac{\partial^2a}{\partial x^2}-\frac{\partial^2a}{\partial y^2}\right)
        \]
        \[
         +\left[\left(\frac{\partial a}{\partial x}\right)^2+\left(\frac{\partial a}{\partial y}\right)^2+
         \left(\frac{\partial a}{\partial z}\right)^2\right] - \left(\frac{\partial b}{\partial y}\right)
         \left(\frac{\partial k}{\partial x}\right)+\left(\frac{\partial b}{\partial z}\right)^2
        \]
        \begin{equation}
        -\left[\left(\frac{\partial a}{\partial t}\right)^2+\left(\frac{\partial b}{\partial t}\right)^2\right]+
        \frac{1}{2}\left(\frac{\partial k}{\partial y}\right)^2
        \end{equation}\\[20pt]

    Finally for \(h_{00} = k\) we get ---

        \[
         \Box k = -a\left(\frac{\partial^2k}{\partial x^2}-\frac{\partial^2k}{\partial y^2}\right) -
         2b\frac{\partial^2k}{\partial x\partial y} + \left(\frac{\partial k}{\partial x}\right)^2+ \left(\frac{\partial k}{\partial y}\right)^2
        \]
        \begin{equation}
        - 2\left(a\frac{\partial^2a}{\partial t^2}+b\frac{\partial^2b}{\partial t^2}\right)-\left[\left(\frac{\partial a}{\partial t}\right)^2+
        \left(\frac{\partial b}{\partial t}\right)^2\right]
        \end{equation}\\[50pt]

    \section{Solution}

    We can find a solution by assuming that \(k = constant\) and \(a = Au(z,t)\) and \(b = Bu(z,t)\) where \(A,B = constants\).
    For this case the most of the equations identically vanish and we are left with --

        \[
         \Box u = -k \frac{\partial^2u}{\partial t^2}\hspace{20pt};\hspace{20pt}\left(\frac{\partial u}{\partial t}\right)^2 =
         \left(\frac{\partial u}{\partial z}\right)^2
        \]
        \[
        u\frac{\partial^2u}{\partial t^2} = -\frac{1}{2}\left(\frac{\partial u}{\partial t}\right)^2\hspace{10pt};
        \hspace{10pt} u\frac{\partial^2u}{\partial z^2} = -\frac{1}{2}\left(\frac{\partial u}{\partial z}\right)^2\hspace{10pt};
        \hspace{10pt}u\frac{\partial^2 u}{\partial t\partial z}=-\frac{1}{2}\left(\frac{\partial u}{\partial t}\right)
        \left(\frac{\partial u}{\partial z}\right)
        \]\\[20pt]

    The second equation from these has two independent solutions --

        \[
         u(z,t) = u_{+}(z+t)\hspace{10pt};\hspace{10pt}u(z,t) = u_{-}(z-t)
        \]

    Using these the LHS of the first equation is identically zero giving \(k = 0\) and the other three equations all reduce to --

        \begin{equation}\label{finaleqn}
        uu'' + \frac{1}{2}{u'}^2 = 0
        \end{equation}\\

    where the prime `` \('\) '' denotes differentiation by \(\lambda_{\pm} = z\pm t\) for \(u_{\pm}\) respectively.\\[10pt]

    To solve this we write \(u' = v\) and thus \(u'' = v\frac{dv}{du}\) and get --\\

        \[
         uv\frac{dv}{du}+\frac{1}{2}v^2 = 0
        \]

    Assuming the \(v \neq 0\) --
        \[
         u\frac{dv}{du} = -\frac{1}{2}v
        \]

    which is readily integrated to give --

        \[
         v = Cu^{-1/2}
        \]

    where C is the integration constant.\\[10pt]

    Further using \(v = \frac{du}{d\lambda}\) we get the solution as ---

        \begin{equation}\label{solution}
        u_{\pm}(z,t) = C\left(z \pm t + \delta\right)^{2/3}
    \end{equation}

    This can be written in the spherical polar coordinates as --

    \begin{equation}
    u_{\pm}(t,r,\theta ,\phi) = C\left(r cos\theta \pm t + \delta\right)^{2/3}
    \end{equation}

    As any other function this can be expanded in terms of the
    spherical harmonics to give the various multipole moments for
    this wave.

    \[
     u_{\pm}(t,r,\theta ,\phi) = \sum_{l=0}^{\infty} \sum_{m=-l}^{+l}C_{l}j_l\left(kr\right)Y_l^m\left(\theta , \phi\right)
    \]

    where the \(j_l(x)\) represents the spherical Bessel functions
    of order \(l\) and \(k = \frac{2 \pi}{\lambda}\) is the wave number
    of the gravitational perturbation.

    This analysis assumes the existence of a flat background
    geometry. This can be assumed to be valid in a FRW expanding
    universe provided that the wavelength of the gravitational waves is small
    enough (equivalently large \(k\)). Thus the above analysis
    represents the large \(k\) limit of the full-calculation.

    To do the full calculation in the FRW background a computational
    code was written in Mathematica and the Ricci tensor computed
    for a second order perturbation in FRW space-time.

\newpage

    \section{Fitting WMAP data}

    To model the anisotropy in the CMBR we can expand the
    fluctuations in terms of the orthogonal spherical harmonic
    functions as --

    \[
    \sum_{l=0}^{\infty}\sum_{m=-l}^{+l}C_{l}Y_l^m\left(\theta , \phi\right)
    \]

    where the \(Y_l^m\) can be written in terms of the Associated
    Legendre Polynomials as --

    \[
    Y_l^m\left(\theta , \phi\right) =
    (-1)^{\frac{m+|m|}{2}}\sqrt{\frac{2l+1}{4\pi}\frac{(l-|m|)!}{l+|m|}!}P_l^m\left(cos
    \theta\right)e^{\iota m\phi}
    \]

    To get the coefficients \(C_{l}\) we can sample the WMAP plot
    at various points of interest and generate a fit to the curve.

    One such fit produced is --

    \[
    l(l+1)|C_l|^2 = 13000 exp(-\frac{l}{300}) J_4(\frac{l}{40})+2500
    \]

    where \(J_4(x)\) represents the Bessel function of order 4.

    Thus the form of \(C_l\) obtained is --

    \[
    C_l = \left[\frac{13000 exp(-\frac{l}{300})
    J_4(\frac{l}{40})+2500}{l(l+1)}\right]^{1/2}
    \]

    Using this form for \(C_l\) the anisotropy can be plotted as 3D
    polar plot to get a picture of the various intricate patterns in
    the gravitational wave spectrum.

    Such a plot was made for a sample of 8 points taken from the
    WMAP plot.

    \begin{table}
    \centering
    \begin{tabular}{c c c}
    \hline\hline
    l & l(l+1)\(|C_l|^2\) & \(C_l\)\\
    \hline
    10 & 900 & 2.861 \\
    100 & 2500 & 0.497\\
    200 & 5800 &0.380\\
    300 & 3500 & 0.197\\
    420 & 1800 & 0.101\\
    540 & 2500 & 0.093\\
    700 & 1800 & 0.004\\
    820 & 2000 & 0.054\\
    \hline
    \end{tabular}
    \label{data_table}
    \end{table}

    This represents only a coarse-grain model. For an accurate model
    more number of data points would have to be sampled resulting in
    a better fit to actual data.

    Though a linear analysis would have given a ripple spectrum but
    there is an explicit mechanism to generate higher-l modes in a
    second order calculation.

    The nonlinear coupling, through product of first derivatives, in the gravitational perturbation can
    result in generation of higher harmonic modes in the
    perturbation. Beginning with lower modes \(l =0 ,1,2\) one can
    generate the observed anisotropy caused due to the nonlinear
    coupling of the wave as seen in the nonlinear equation.

    This nonlinear coupling can transfer energy from lower l-modes to
    higher l-modes. This results in an anisotropic wrinkled
    space-time. This can be linked to the density perturbations as
    observed in CMBR and WMAP as done below.

\section{Kinematics of distribution of matter through a
        simple
        model}
\subsection{Diffusion equation model}
In our context the word diffusion applies to motion of particles of
matter {matter may be normal matter,dark matter or dark energy}. Let
us try to solve the diffusion problem in one dimensions. {the method
for higher dimensions follows similarly}Divide the space into thin
parts.The process of diffusion can be considered effectively as the
problem of random walk which deals with the statistical progress of
a particle equally likely to move forwards or backwards. At any
instant of time t half the particles in a part at the position x
diffuse to the part at \(x+\Delta x\) and half the particles diffuse
to part at \(x-\Delta x\).The variables x and t being continuous
require a population density function n(x,t);its meaning is that
n(x,t)\(\Delta x\)is the population occupying a short section
\(\Delta x\). Hence we get
\begin{displaymath}
\frac{1}{2}n(x+\Delta x,t)+\frac{1}{2}n(x-\Delta x,t)=n(x,t +\Delta
t)
\end{displaymath}
Subtracting n(x,t) from both sides, it is supposed that the
difference is very small compared to n(x,t) itself, we get
\begin{displaymath}
\Rightarrow \frac{1}{2}[n(x+\Delta
x,t)-n(x,t)]-\frac{1}{2}[n(x,t)-n(x-\Delta x,t)]=n(x,t +\Delta
t)-n(x,t)
\end{displaymath}
\begin{displaymath}
\Rightarrow \frac{1}{2}\left(\left(\frac{\partial n}{\partial
x}\right)_{x}-\left(\frac{\partial n}{\partial x}\right)_{x-\Delta
x}\right)\Delta x=\left(\frac{\partial n}{\partial t}\right)\Delta t
\end{displaymath}
\begin{displaymath}
\Rightarrow \frac{1}{2}(\Delta x)^{2}\frac{\partial^{2}n}{\partial
x^{2}}=\left(\frac{\partial n}{\partial t}\right)\Delta t
\end{displaymath}
The final step is to assume that, with  a large number of particles,
\(\Delta x\) and \(\Delta t\) can be reduced indefinitely (which
makes the above approximations exact) but keeping the ratio
\(D=\frac{\frac{1}{2}(\Delta x)^{2}}{\Delta t}\) constant. Thus we
have the diffusion equation as
\begin{displaymath}
D\frac{\partial^{2}n}{\partial x^{2}}=\left(\frac{\partial
n}{\partial t}\right)
\end{displaymath}
which means that the random diffusing process which we envisage does
indeed lead to the diffusion equation. A special solution of
diffusion equation is the Gaussian which is
\begin{displaymath}
n(x,t)=(\pi D t)^{\frac{-1}{2}}e^{\left(\frac{-x^{2}}{4Dt}\right)}
\end{displaymath}
The above equation gives the distribution of number of particles
 with respect to position at a given time t. If there was very
distribution at \(x_{0}\) at time \(t_{0}\) then we write the
Gaussian as
\begin{equation}\label{eq diffusion solution}
n(x,t)=(\pi D
(t-t_{0}))^{\frac{-1}{2}}e^{\left(\frac{-(x-x_{0})^{2}}{4D(t-t_{0}}\right)}
\end{equation}
As mentioned in the introduction we are trying to model the
redistribution of matter in the Universe as a process of diffusion
running backwards in time\@. Looked at the scale of the Universe the
various points of accumulation of normal matter correspond to
super-clusters. At the scale of super-clusters points of
accumulation of normal matter correspond to clusters and the scale
of clusters to galaxies. Firstly we explain the clustering and
clumping of normal matter at the scale of the Universe.Similar
arguments are valid at the scale of super-clusters,clusters and
galaxies. Now if the process of clustering of normal matter is
diffusion running backwards in time and equation \ref{eq diffusion
solution} is a solution of diffusion equation and \(t_{0}\) denotes
the present time then \(x_{0}\) would be the position of a typical
super-cluster. But equation \ref{eq diffusion solution} would mean
that all the matter is concentrated at the point \(x_{0}\) at
present time. This would mean that a super-cluster has no size which
is wrong. Therefore we add a \(-\sigma^{2}\) term to equation
\ref{eq diffusion solution} where \(\sigma\) is the standard
deviation which is an estimate of the size of super-cluster. For an
Gaussian \(\sigma \) corresponds to half of the width of the
Gaussian at half maximum which is 65\% of the total width of the
Gaussian. For a Gaussian representing a super-cluster the radius of
super cluster is roughly \(3 \sigma\) as \(3 \sigma=95 \%\) of all
matter is within the cluster entity of this radius.

Now having modeled the distribution of normal matter through a
diffusion equation we can account for the redistribution of
dark-matter and dark-energy also in a similar way. The most
important point to be noted is that the Gaussian for different forms
of matter will have different values of Diffusion coefficient D and
different values for standard deviation \(\sigma\). This is the main
reason why different forms of matter differentiate among each other
and ended up at different places in the Universe.  1-d as well as
2-d plots were made using Matlab of density of normal matter v/s
position for various times at the scale of Universe. From these
figures one can see how a uniform distribution of normal matter
ultimately led to the formation of super-clusters with sharp
densities. The formation of clusters within a super cluster and of
galaxies within a cluster can be explained similarly. Note that in
the real diffusion process of a gas the series of pictures would be
from sharp to wide gaussians. The radius of the Universe is \(4000
mega-parsecs \sim 10^{26}m\).In the 1-d plot 1 unit=10 mega-parsecs
and in 2-d plot 1 unit=100 mega-parsecs were used.

\newpage
We now draw a rough analogy between the distribution of various
forms of matter in the Universe and condensation of air. The rough
mass composition of air in our atmosphere is 70\% of \(N_{2}\) 27\%
of \(O_{2}\),1\% of \(CO_{2}\),0.5\% of \(H_{2}O\) and 0.1 \% of CO.
As the temperature keeps dropping different gases start to condense.
The process begins with the condensation of water-vapor. At various
points in air one finds formation of droplets of water at different
places as we go closer to \(0^{\tiny{0}}C\). Condensation is nothing
but bringing water molecules which were far closer to each other
until sufficient accumulation takes place as the phase changes from
gas to liquid. As the temperature goes down further to
\(-40^{\tiny{0}}C\) the \(CO_{2}\) in the atmosphere also condenses
and similarly various gases condense at various stages. Now the
redistribution of various forms of matter in the Universe is
analogous to condensation of air in the sense that various forms of
matter accumulate at different points in space.Note that the mass
composition of our Universe is 69.7475\% of Dark Energy present in
voids,29.2475\% of Dark matter out which 2.5\% is in galaxy halos
and the rest is in voids,0.5\% of normal matter and .005 \% of
radiation present everywhere in the Universe.

\begin{tabular}{|p{1in}|p{1in}|p{1in}|p{1in}|} \hline

    Name&Number&Size&Mass \\ \hline
    Universe&1&order of \(10^{26}\) meters (around 4000 mega parsecs  \(10^{9}\) light years&\(10^{20}\) solar masses \\ \hline
    Supercluster &roughly 1000& 400 mega parsecs &\(10^{17}\) solar masses \\ \hline
    Cluster&256000&40  mega parsecs&\(10^{15}\) solar masses \footnote{source is Wikipedia}   \\ \hline
    Galaxy&\(10^{11}\) galaxies in the Universe&4 mega parsecs & \(10^{9}\) solar masses  \\ \cline{2-2}
    &&& \\ \hline
\end{tabular}

Now we try to establish a criterion to find out whether or not a
given set of cluster of galaxies form a super-cluster.Let the size
of each grid be 1 unit \(\times\) 1 unit\@.Now from the figure one
can make out various patterns of super-cluster of cluster of
galaxies\@.But can we try to obtain a quantitative result in order
to define a super-cluster\@.For this purpose consider any pattern
from the figure\@.For example we consider the pattern of clusters
with coordinates:

(28,30);(27,32);(30,32);(26,34);(29,34);(31,36);(29,37);(31,38);(29,40)

Since different clusters are separated from each other by different
distances we need to obtain a quantity which represents the average
value of all distances\@.Now if the evaluated mean is less than some
fixed value then the given group of clusters form a super
cluster\@.If (\(x_{i},y_{i}\)) and (\(x_{j},y_{j}\)) are the
coordinates of the ith and jth  cluster then the following
quantities may be evaluated for a set of n clusters :
\begin{enumerate}
\item
\begin{displaymath}
\left(\frac{\sum_{i=1}^{n}\sum_{j=i+1}^{n}[(x_{i}-x_{j})^{2}+(y_{i}-y_{j})^{2}]}{n}\right)^{\frac{1}{2}}=10.203
\end{displaymath}
\item
\begin{displaymath}
\left(\frac{\sum_{i=1}^{n}\sum_{j=i+1}^{n}[(x_{i}-x_{j})^{2}+(y_{i}-y_{j})^{2}]}{^{n}C_{2}}\right)^{\frac{1}{2}}=5.102
\end{displaymath}
\item
\begin{displaymath}
\left(\frac{\sum_{i=1}^{n}\sum_{j=i+1}^{n}[(x_{i}-x_{j})^{2}+(y_{i}-y_{j})^{2}]^{\frac{1}{2}}}{n}\right)=20.403
\end{displaymath}
\item
\begin{displaymath}
\left(\frac{\sum_{i=1}^{n}\sum_{j=i+1}^{n}[(x_{i}-x_{j})^{2}+(y_{i}-y_{j})^{2}]^{\frac{1}{2}}}{^{n}C_{2}}\right)=5.101
\end{displaymath}
\end{enumerate}

\begin{tabular}{|c|c|p{.7in}|p{.7in}|c|c|} \hline
Sl.no&positions of cluster&\(\Delta\)x i.e, x separation&\(\Delta\)y
i.e, y separation&\({(\Delta x)}^2+{(\Delta y)}^2\)&\({({(\Delta
x)}^2+{(\Delta y)}^2)}^\frac{1}{2}\)\\ \hline
1&(28,30)(27,32)&1&2&5&\(\sqrt{5}\) \\ \hline
2&(28,30)(30,32)&2&2&8&\(2\sqrt{2}\)\\ \hline
3&(28,30)(26,34)&2&4&20&\(2\sqrt{5}\)\\ \hline
4&(28,30)(29,34)&1&4&17&\(\sqrt{17}\)\\ \hline
5&(28,30)(31,36)&3&6&45&\(3\sqrt{5}\) \\ \hline
6&(28,30)(29,37)&1&7&50&\(5\sqrt{2}\) \\ \hline
7&(28,30)(31,38)&3&8&73&\(\sqrt{73}\) \\ \hline
8&(28,30)(29,40)&1&10&101&\(\sqrt{101}\)\\ \hline
9&(27,32)(30,32)&3&0&9&3\\ \hline
10&(27,32)(26,34)&1&2&5&\(2\sqrt{5}\)\\ \hline
11&(27,32)(29,34)&4&4&8&\(2\sqrt{2}\)\\ \hline
12&(27,32)(31,36)&4&4&32&\(4\sqrt{2}\) \\ \hline
13&(27,32)(29,37)&2&5&29&\(\sqrt{29}\) \\ \hline
14&(27,32)(31,38)&4&6&52&\(\sqrt{52}\) \\ \hline
15&(27,32)(29,40)&2&8&68&\(2\sqrt{17}\)\\ \hline
16&(30,32)(26,34)&4&2&20&\(2\sqrt{5}\)\\ \hline
17&(30,32)(29,34)&1&2&5&\(\sqrt{5}\)\\ \hline
18&(30,32)(31,36)&1&4&17&\(\sqrt{17}\) \\ \hline
19&(30,32)(29,37)&1&5&26&\(\sqrt{26}\) \\ \hline
20&(30,32)(31,38)&1&6&37&\(\sqrt{37}\) \\ \hline
21&(30,32)(29,40)&1&8&65&\(2\sqrt{65}\)\\ \hline
22&(26,34)(29,34)&3&0&9&3 \\ \hline
23&(26,34)(31,36)&2&2&8&\(2\sqrt{2}\) \\ \hline
24&(26,34)(29,37)&3&3&18&\(3\sqrt{2}\) \\ \hline
25&(26,34)(31,38)&5&4&41&\(2\sqrt{41}\) \\ \hline
26&(26,34)(29,40)&3&6&45&\(3\sqrt{5}\)\\ \hline
27&(29,34)(31,36)&2&2&8&\(2\sqrt{2}\) \\ \hline
28&(29,34)(29,37)&0&3&9&3 \\ \hline
29&(29,34)(31,38)&2&4&20&\(2\sqrt{5}\) \\ \hline
30&(29,34)(29,40)&0&6&36&\(6\)\\ \hline
31&(31,36)(29,37)&2&1&5&\(\sqrt{5}\) \\ \hline
32&(31,36)(31,38)&0&2&4&2 \\ \hline
33&(31,36)(29,40)&2&4&20&\(2\sqrt{5}\)\\ \hline
34&(29,37)(31,38)&2&1&5&\(\sqrt{5}\) \\ \hline
35&(29,37)(29,40)&0&3&9&3\\ \hline
36&(31,38)(29,40)&2&2&8&\(2\sqrt{2}\)\\ \hline

\end{tabular}

\section{Density perturbation using FRW metric model}

We discuss the model based on the Friedman-Robertson-Walker (FRW)
model. The FRW metric is given by
\begin{displaymath}
ds^{2}=c^{2}dt^{2}-a^{2}(t)(dr^{2}+r^{2}(d \theta)^{2}+r^{2}\sin
^{2} \theta (d \phi)^{2})
\end{displaymath}
Einstein's field equation :
\begin{displaymath}
R_{\mu \nu}-\frac{1}{2}g_{\mu \nu}R=\left(\frac{8 \pi G
}{c^{4}}\right)T_{\mu \nu}
\end{displaymath}
where \(R_{\mu \nu}\) is the Ricci tensor, \(g_{\mu \nu}\) is the
Einstein tensor,R is the Ricci scalar and \(T_{\mu \nu}\) is the
stress-energy tensor.With the above two equations get at the
following two equations \footnote{Reference 1 Chapter 4}
\begin{displaymath}
\frac{2\ddot{a}}{a}+\frac{\dot{a^{2}}+kc^{2}}{a^{2}}=\left(\frac{8
\pi G  }{c^{4}}\right)T^{3}_{3}
\end{displaymath}
\begin{displaymath}
\frac{\dot{a^{2}}+kc^{2}}{a^{2}}=\left(\frac{8 \pi G \rho_{0}
}{3c^{2}}\right)T^{0}_{0}
\end{displaymath}
where k is the curvature constant.

Assuming that the Universe is filled with Pressure less dust then
the stress energy tensor \(T_{\mu \nu}\) becomes
\begin{displaymath}
\left(\begin{array}{cccc}
\rho & 0 & 0& 0 \\
 0 & 0& 0 & 0 \\
 0 & 0& 0 & 0 \\
 0 & 0& 0 & 0
\end{array}\right)
\end{displaymath}
Hence assuming a pressureless dust-dominated Universe and taking the
curvature constant to be 0 in the above two equations we get the
following two equations.
\begin{equation}\label{eq a}
\frac{2\ddot{a}}{a}+\frac{\dot{a^{2}}}{a^{2}}=0
\end{equation}
\begin{equation}\label{eq a rho}
\frac{\dot{a^{2}}}{a^{2}}=\left(\frac{8 \pi G \rho_{0}
}{3c^{4}}\right)\frac{a_{0}^{3}}{a^{3}}
\end{equation}
Solving equation \ref{eq a rho} we get
\begin{displaymath}
\dot{a^{2}}=\left(\frac{8 \pi G \rho_{0}
}{3c^{4}}\right)\frac{a_{0}^{3}}{a}
\end{displaymath}
If \(t_{0}\) represents the present time then
\begin{displaymath}
\left(\frac{\dot{a}}{a}\right)_{t_{0}}=H_{0}
\end{displaymath}
\begin{equation}\label{eq Ho}
\rho_{0}\equiv\frac{3H_{0}^{2}c^{4}}{8 \pi G }
\end{equation}
We now solve equation \ref{eq a rho} by rewriting as
\begin{displaymath}
\dot{a^{2}}=H_{0}^{2}\frac{a_{0}^{3}}{a}
\end{displaymath}
\begin{displaymath}
\Rightarrow \dot{a}=H_{0}
\frac{a_{0}^{\frac{3}{2}}}{a^{\frac{1}{2}}}
\end{displaymath}
\begin{displaymath}
\Rightarrow \int {a^{\frac{1}{2}}}\,da =\int H_{0}
a_{0}^{\frac{3}{2}}\,dt
\end{displaymath}

\begin{displaymath}
\Rightarrow {a^{\frac{3}{2}}} =\frac{3}{2} H_{0}
a_{0}^{\frac{3}{2}}t
\end{displaymath}
Assuming a=0 at t=0 the arbitrary constant that arises out of
integration is set to 0.

Putting
\begin{equation}\label{eq to}
H_{0}= \frac{2}{3 t_{0}}
\end{equation}
\begin{displaymath}
\Rightarrow {a^{\frac{3}{2}}} =\frac{3}{2} \frac{2}{3 t_{0}}
a_{0}^{\frac{3}{2}}t
\end{displaymath}
Putting \(H_{0}= \frac{2}{3 t_{0}}\) where \(t_{0}\) is the present
time.
\begin{equation}\label{eq sol a}
\Rightarrow a = a_{0} \left(\frac{t}{t_{0}}\right) ^{\frac{2}{3}}
\end{equation}
One may as well check that this solution of a satisfies \ref{eq a}
by substituting
\(\dot{a}=\frac{2}{3}a_{0}\left(\frac{t}{t_{0}}\right)
^{\frac{-1}{3}}\frac{1}{t_{0}}\) \ and \
\(\ddot{a}=\frac{-2}{9}a_{0}\left(\frac{t}{t_{0}}\right)
^{\frac{-4}{3}}(\frac{1}{t_{0}})^{2}\)

 As we are considering a model of Universe in which mass (M(t))
is conserved we have
\begin{displaymath}
\left(\frac{d(M(t))}{dt}\right)=0
\end{displaymath}
\begin{displaymath}
\Rightarrow \left(\frac{d(\rho(t)V(t))}{dt}\right)=0
\end{displaymath}
where \(\rho(t)\)and\(V(t)\) are the density and volume of the
Universe respectively.
\begin{displaymath}
\Rightarrow
\rho(t)\left(\frac{d(V(t))}{dt}\right)+V(t)\left(\frac{d(\rho(t))}{dt}\right)=0
\end{displaymath}
 \(V(t)=bR^{3}\) where R is the radius of the Universe which is a
 function of time and b is a constant. Put \(\rho(t)\equiv \rho\)
 then
\begin{displaymath}
3b\rho
R^{2}\left(\frac{d(R)}{dt}\right)+bR^{3}\left(\frac{d(\rho)}{dt}\right)=0
\end{displaymath}
If \(\frac{d(R)}{dt} \equiv \dot{R}\)\ and\ \(\frac{d(\rho)}{dt}
\equiv \dot{\rho}\) then
\begin{displaymath}
3 \rho \dot{R}=-R \dot{\rho}
\end{displaymath}
Since \(\frac{\dot{R}}{R}=\frac{\dot{a}}{a}\)
\begin{displaymath}
 \frac{\dot{\rho}}{\rho}=\frac{-3\dot{R}}{R}=\frac{-3\dot{a}}{a}
\end{displaymath}
Finally
\begin{equation}\label{eq rho}
\frac{\dot{\rho}}{\rho}=\frac{-3\dot{a}}{a}
\end{equation}
\begin{displaymath}
 \frac{\dot{\rho}}{\rho}=\frac{-3\dot{R}}{R}=\frac{-3\dot{a}}{a}
\end{displaymath}
Conservation of mass would imply that
\begin{displaymath}
\rho R^{3}=\rho_{0} R_{0}^{3}
\end{displaymath}
Using
\begin{displaymath}
\frac{a}{a_{0}}=\frac{R}{R_{0}}
\end{displaymath}
\begin{displaymath}
\rho \left(\frac{a}{a_{0}}\right)^{3}=\rho_{0}
\end{displaymath}
Substituting 'a' as a function of time from equation \ref{eq sol a}
we get
\begin{equation}\label{eq sol rho}
\rho=\frac{\rho_{0}t_{0}^{2}}{t^{2}}
\end{equation}

Till now we have discussed equations and solutions which are valid
provided our Universe is isotropic with respect to space i.e, if the
scale factor 'a' were function of time alone. The FRW model is
conventionally done in spherical-polar co-ordinates. Now if isotropy
and homogeneity is broken we modify a(t) so that \(\frac{\partial
a}{\partial \theta}\neq 0\) and  \(\frac{\partial a}{\partial
\phi}\neq 0\).For a 3 sphere we write down \(a(t,\theta,\phi)\).
\begin{equation}\label{eq a change}
a(t,\theta,\phi)=a_{1}(t)+a_{2}(t, \theta ,\phi)
\end{equation}
Similarly for the matter part we write down \(\rho(t, \theta ,
\phi)\)
\begin{equation}\label{eq rho change}
\rho(t,\theta,\phi)=\rho_{1}(t)+\rho_{2}(t, \theta ,\phi)
\end{equation}
If P represents the pressure and T the temperature at WMAP time we
have \( \frac{\Delta \rho}{\rho}\simeq \frac{\Delta T}{T} \simeq
\frac{\Delta P}{P}\simeq 10^{-5} \) which increases with subsequent
time. Hence \( \frac{\rho_{2}}{\rho}\simeq \frac{a_{2}}{a} \simeq
10^{-5}\) at WMAP time (\(10^{13} seconds\).So in the equations
\ref{eq a},\ref{eq a rho} and \ref{eq rho} we include the effect of
\(a_{2}\) and try to obtain the new set of equations with
\(\frac{a_{2}}{a} \ll 1 \)\ , \ \(\frac{\rho_{2}}{\rho} \ll 1\).

The detailed structure formation on the Universe as seen at z
(doppler shift) =0 to 1 with voids,cluster of galaxies and various
kinds of fluctuations requires super computing models to put
observed data from HST, SDSS etc which is outside the scope of this
project. We solve for density fluctuations from WMAP i.e,z=1000 till
z=5 and density fluctuations from z=5 to z=1. Recalling that
\begin{displaymath}
1+z=\frac{a(t_{0})}{a(t)}=\left(\frac{t_{0}}{t}\right)^{\frac{2}{3}}
\end{displaymath}
Now combining \ref{eq a} and \ref{eq a rho} we get
\begin{equation}\label{eq net a rho}
\frac{2\ddot{a}}{a}+\left(\frac{8 \pi G \rho_{0}
}{3c^{4}}\right)\frac{a_{0}^{3}}{a^{3}}=0
\end{equation}
Introducing inhomogeneity and anisotropy in \(a \equiv a(t , \theta
, \phi)\) as explained above we have
\begin{displaymath}
\frac{2(\ddot{a_{1}}+\ddot{a_{2}})}{a_{1}+a_{2}}+ \frac{8 \pi G
\rho_{0} a_{0}^{3}}{3c^{4}(a_{1}+a_{2})^{3}}=0
\end{displaymath}
Note that \(a_{1} \equiv a_{1}(t)\) and \(a_{2} \equiv a_{2}(t
,\theta , \phi)\)
\begin{displaymath}
\Rightarrow
\frac{2(\ddot{a_{1}}+\ddot{a_{2}})}{a_{1}}\left(1+\frac{a_{2}}{a_{1}}\right)^{-1}+
\frac{8 \pi G \rho_{0} a_{0}^{3}}{3c^{4}a_{1}^{3}}\left(1+
\frac{a_{2}}{a_{1}}\right)^{-3}=0
\end{displaymath}
In the binomial expansion
\(\left(1+\frac{a_{2}}{a_{1}}\right)^{-1}\) and
\(\left(1+\frac{a_{2}}{a_{1}}\right)^{-3}\) we neglect the terms
having the power of \(a_{2}\) greater than 1
\begin{displaymath}
\Rightarrow
\frac{2(\ddot{a_{1}}+\ddot{a_{2}})}{a_{1}}\left(1-\frac{a_{2}}{a_{1}}\right)+
\frac{8 \pi G \rho_{0} a_{0}^{3}}{3c^{4}a_{1}^{3}}\left(1-
\frac{3a_{2}}{a_{1}}\right)=0
\end{displaymath}

\begin{displaymath}
Approximating \ \frac{2\ddot{a_{2}}a_{2}}{a_{1}a_{1}}\  to \  0 \ we
\ get
\end{displaymath}

\begin{displaymath}
\frac{2\ddot{a_{1}}}{a_{1}}-\frac{2\ddot{a_{1}}a_{2}}{a_{1}a_{1}}+\frac{2\ddot{a_{2}}}{a_{1}}+\frac{8
\pi G \rho_{0}
a_{0}^{3}}{3c^{4}a_{1}^{3}}-\frac{3a_{2}}{a_{1}}\left(\frac{8 \pi G
\rho_{0} a_{0}^{3}}{3c^{4}a_{1}^{3}}\right)=0
\end{displaymath}
Using equation \ref{eq net a rho} we have
\begin{displaymath}
\frac{2\ddot{a_{1}}}{a_{1}}+\left(\frac{8 \pi G \rho_{0}
a_{0}^{3}}{3c^{4}a_{1}^{3}}\right)\frac{a_{0}^{3}}{a_{1}^{3}}=0
\end{displaymath}
Therefore
\begin{equation}\label{eq a2}
\ddot{a_{2}}-\frac{\ddot{a_{1}}a_{2}}{a_{1}}- \frac{4 \pi G
\rho_{0}}{c^{4}}\left(\frac{a_{0}^{3}}{a_{1}^{3}}\right)a_{2}=0
\end{equation}
Now since \(a_{1}\) plays role of a in equation \ref{eq net a rho}
and using equation \ref{eq Ho} we get
\begin{displaymath}
\frac{\ddot{a_{1}}}{a_{1}}=\frac{-H_{0}^{2}}{2}\left(\frac{a_{0}^{3}}{a_{1}^{3}}\right)
\end{displaymath}
Using the above expression and equation \ref{eq Ho} equation \ref{eq
a2} can be written as
\begin{displaymath}
\ddot{a_{2}}+\frac{H_{0}^{2}}{2}\left(\frac{a_{0}^{3}}{a_{1}^{3}}\right)a_{2}-\frac{3
H_{0}^{2}}{2}\left(\frac{a_{0}^{3}}{a_{1}^{3}}\right)a_{2}=0
\end{displaymath}
\begin{displaymath}
\Rightarrow
\ddot{a_{2}}-H_{0}^{2}\left(\frac{a_{0}^{3}}{a_{1}^{3}}\right)a_{2}=0
\end{displaymath}
Since \(a_{1}\) plays the role of 'a' in equation \ref{eq sol a} we
know \(a_{1}\) as a function of time and using equation \ref{eq to}
to substitute for\(H_{0}\) we get
\begin{equation}\label{eq a2,t}
\ddot{a_{2}}-\frac{4 a_{2}}{9 t^{2}}=0
\end{equation}
Plotting a graph of \(a_{2}\) v/s time gives (\(a_{2}^{1} \equiv
\dot{a_2})\)

Introducing inhomogeneity and anisotropy in \(\rho \equiv \rho (t ,
\theta , \phi)\) as explained above we have
\begin{displaymath}
\dot{\rho_{1}}+\dot{\rho_{2}}=\frac{-3(\dot{a_{1}}+\dot{a_{2}})(\rho_{1}+\rho_{2})}{a_{1}+a_{2}}
\end{displaymath}
Note that \(\rho_{1} \equiv \rho_{1}(t)\) and \(\rho_{2} \equiv
\rho_{2}(t ,\theta , \phi)\)
\begin{displaymath}
\Rightarrow
\dot{\rho_{1}}+\dot{\rho_{2}}=\frac{-3(\dot{a_{1}}+\dot{a_{2}})(\rho_{1}+\rho_{2})}{a_{1}}\left(1+\frac{a_{2}}{a_{1}}\right)^{-1}
\end{displaymath}
In the binomial expansion
\(\left(1+\frac{a_{2}}{a_{1}}\right)^{-1}\)  we neglect the terms
having the power of \(a_{2}\) greater than 1
\begin{displaymath}
\Rightarrow
\dot{\rho_{1}}+\dot{\rho_{2}}=\frac{-3(\dot{a_{1}}+\dot{a_{2}})(\rho_{1}+\rho_{2})}{a_{1}}\left(1-\frac{a_{2}}{a_{1}}\right)
\end{displaymath}
\begin{displaymath}
Approximating \
 \frac{-3\dot{a_{1}}\rho_{2}}{a_{1}}\left(\frac{-a_{2}}{a_{1}}\right)
\ , \  \frac{-3\dot{a_{2}}\rho_{2}}{a_{1}}  \ , \
 \frac{-3\dot{a_{2}}\rho_{1}}{a_{1}}\left(\frac{-a_{2}}{a_{1}}\right)
\ and \
\frac{-3\dot{a_{2}}\rho_{1}}{a_{1}}\left(\frac{-a_{2}}{a_{1}}\right)
 \  to \  0 \ we \ get
\end{displaymath}
\begin{displaymath}
\dot{\rho_{1}}+\dot{\rho_{2}}=
-\frac{3\dot{a_{1}}\rho_{1}}{a_{1}}-\frac{3\dot{a_{1}}\rho_{2}}{a_{1}}-
\frac{3\dot{a_{1}}\rho_{1}}{a_{1}}\left(\frac{-a_{2}}{a_{1}}\right)
-\frac{3\dot{a_{2}}\rho_{1}}{a_{1}}
\end{displaymath}
Using \ref{eq rho}
\begin{displaymath}
\frac{\dot{\rho_{1}}}{\rho_{1}}=\frac{-3\dot{a_{1}}}{a_{1}}
\end{displaymath}
Therefore
\begin{equation}\label{eq rho2}
\dot{\rho_{2}}= -\frac{3\dot{a_{1}}\rho_{2}}{a_{1}}+
\frac{3\dot{a_{1}}\rho_{1}}{a_{1}}\left(\frac{a_{2}}{a_{1}}\right)
-\frac{3\dot{a_{2}}\rho_{1}}{a_{1}}
\end{equation}

We have simulated the process of structure formation in our Universe
as a diffusion process running backwards in time\@.This means that
the density of normal matter satisfies the diffusion
equation\@.Actually the process of structure formation of normal
matter is anisotropic in space.This means that the \(\rho_{2}\) term
of the density satisfies the diffusion equation i.e,
\begin{equation}\label{eq diff rho}
\frac{\partial \rho_{2}}{\partial t}=D \nabla^{2} \rho_{2}
\end{equation}
where D is the diffusion coefficient. But from equation \ref{eq
rho2} \(\rho_{2}\) also satisfies
\begin{displaymath}
\dot{\rho_{2}}= -\frac{3\dot{a_{1}}\rho_{2}}{a_{1}}+
\frac{3\dot{a_{1}}\rho_{1}}{a_{1}}\left(\frac{a_{2}}{a_{1}}\right)
-\frac{3\dot{a_{2}}\rho_{1}}{a_{1}}
\end{displaymath}
Now this is analogous to a diffusion equation since this can be
rearranged as
\begin{displaymath}
\dot{\rho_{2}}+\frac{3\dot{a_{1}}\rho_{2}}{a_{1}}=
\frac{3\dot{a_{1}}\rho_{1}}{a_{1}}\left(\frac{a_{2}}{a_{1}}\right)
-\frac{3\dot{a_{2}}\rho_{1}}{a_{1}}
\end{displaymath}
When analogy is drawn between this equation and the diffusion
equation we see that the L.H.S is analogous to diffusion equation
and the R.H.S suggests the existence of a source within the system
undergoing diffusion\@.For the time being let us assume the source
term to be absent i.e, remove the perturbation in a\@.Now in the
absence of the perturbation term the equation becomes
\begin{equation}\label{eq rho2 no source}
\dot{\rho_{2}}= -\frac{3\dot{a_{1}}\rho_{2}}{a_{1}}
\end{equation}
Compare the above equation with diffusion equation in one dimension
{for simplicity}\@.Therefore from equation \ref{eq diff rho} we get
\begin{displaymath}
\dot{\rho_{2}}=\frac{\partial \rho_{2}}{\partial
t}=D\frac{\partial^{2}\rho_{2}(z)}{\partial z^{2}}
\end{displaymath}
From the above two equations it follows that
\begin{equation}\label{eq diff rho2 no source}
D \frac{\partial^{2}\rho_{2}(z)}{\partial
z^{2}}=-\frac{3\dot{a_{1}}\rho_{2}}{a_{1}}
\end{equation}
This means that D\(\nabla^{2}\) acting on \(\rho_{2}\) is equal
\(\frac{-3\dot{a_{1}}}{ a_{1}}\) times \(\rho_{2}\).Note that
\(\frac{3\dot{a_{1}}}{ a_{1}}\) is a constant with respect to
position .In such cases the variation of density with position is a
sine function and cosine function or a linear combination of both
i.e an exponential function.One must also observe that
\(\frac{3\dot{a_{1}}}{ a_{1}}\) is just 3 times the value of
Hubble's constant at different epochs. In equation  \ref{eq diff
rho} if \(\nabla^{2}\) is expanded in spherical polar co-ordinate
system as
\begin{displaymath}
\nabla^{2} \equiv
\nabla^{2}_{r}+\nabla^{2}_{\theta}+\nabla^{2}_{\phi}
\end{displaymath}
where \(\nabla^{2}_{i} \  ; \  i \ = \ r, \ \theta \ , \ \phi  \)
are in general functions of all the three variables \(r \ , \ \theta
\ , \ \phi \)

If anisotropy and inhomogeneity is taken via \(\rho_{2}(t , \theta ,
\phi)\) and \( a_{2}(t , \theta , \phi)\) then diffusion equation
\ref{eq diff rho} takes the form

\begin{displaymath}
\frac{\partial \rho_{2}}{\partial t}=D
(\nabla^{2}_{r}+\nabla^{2}_{\theta}+\nabla^{2}_{\phi})
\end{displaymath}
From equation \ref{eq rho2 no source} we have
\begin{displaymath}
\dot{\rho_{2}}= -\frac{3\dot{a_{1}}\rho_{2}}{a_{1}}
\end{displaymath}
From the above two equations we get
\begin{displaymath}
D(\nabla^{2}_{r}+\nabla^{2}_{\theta}+\nabla^{2}_{\phi})=-\frac{3\dot{a_{1}}\rho_{2}}{a_{1}}
\end{displaymath}
Now the solution for the above equation will be of the form
\begin{displaymath}
\rho_{2}(t , r, \theta , \phi)=\sum_{l}c_{l}J_{l}(kr)Y_{lm}(\theta ,
\phi)
\end{displaymath}
 We consider the sampling of data of CMBR improving in resolution
from COBE to WMAP to Planck satellites will give better angular
correlations in \(\theta\) and \(\phi\). So that \(a_{2}\) and
\(\rho_{2}\) the anisotropic perturbations of space-time geometry
and matter density can be given values at neighboring \(\theta\) and
\(\theta \pm d\theta\) and the difference equation converted to a
differential equation.Hence a \(\nabla^{2}\) acts on \(\rho_{2}\) in
equation \ref{eq diff rho}.This feature is repeated at scale of
super-cluster,cluster and galaxy.

Consider equation \ref{eq rho2} and substituting \(a_{1}\),\(\dot
a_{1}\) using equation \ref{eq sol a} and substituting for \(rho\)
from equation \ref{eq sol rho} we arrive at the equation
\begin{equation}\label{eq rho2 t}
\dot{\rho_{2}}=\frac{-2
\rho_{2}}{t}+\frac{\rho_{0}}{a_{0}}\left(\frac{2a_{2}}{t}-3\dot{a_{2}}\right)\left(\frac{t_{0}}{t}\right)^\frac{8}{3}
\end{equation}
Considering the equation \ref{eq rho2 no source} substituting
\(a_{1}\),\(\dot a_{1}\) using equation \ref{eq sol a} we get
\begin{equation}\label{eq rho2 no source t}
\dot{\rho_{2}}=\frac{-2 \rho_{2}}{t}
\end{equation}
This would be the case if in equation \ref{eq rho2 t}
\begin{equation}
\frac{2a_{2}}{t}=3\dot{a_{2}}
\end{equation}
\begin{equation}
\Rightarrow \frac{\dot{a_{2}}}{a_{2}}=\frac{2}{3t}
\end{equation}
From  equation \ref{eq to} one can conclude that
\begin{displaymath}
H= \frac{2}{3
t}=\frac{\dot{a}}{a}=\frac{\dot{a_{1}}+\dot{a_{2}}}{a_{1}+a_{2}}
\end{displaymath}
From the above two equations it follows that the anisotropic part of
Hubble's constant scales in the same way as Hubble when diffusion
without a source takes place.This is exactly the situation during
WMAP when clustering has just started. From equation \ref{eq rho2 t}
\begin{displaymath}
\dot{\rho_{2}}=\frac{-2
\rho_{2}}{t}+\frac{\rho_{0}}{a_{0}}\left(\frac{2a_{2}}{t}-3\dot{a_{2}}\right)\left(\frac{t_{0}}{t}\right)^\frac{8}{3}
\end{displaymath}
If \(\dot{a_{2}}\) is approximated to 0 in the equation \ref{eq rho2
t} which is the present state of Universe (from 1 billion years
onwards) where in the change in perturbation is made very small and
using equation \ref{eq sol a} and equation \ref{eq sol rho} in the
equation \ref{eq rho2 t} we get
\begin{displaymath}
\dot{\rho_{2}}+\frac{2
\rho_{2}}{t}=\frac{\rho_{1}}{a_{1}}\left(\frac{2a_{2}}{t}\right)
\end{displaymath}
This is an equation in which R.H.S contains the anisotropic part
\(a_{2}\) and roughly models the present state of our Universe.

Considering the equation \ref{eq diff rho2 no source} substituting
\(a_{1}\),\(\dot a_{1}\) using equation \ref{eq sol a} we get
\begin{equation}\label{eq diff rho2 no source t}
D \frac{\partial^{2}\rho_{2}(z)}{\partial z^{2}}=\frac{-2
\rho_{2}}{t}
\end{equation}
By inspection
\(\rho_{20}exp\left(\frac{-ix}{\left(\frac{Dt}{2}\right)^\frac{1}{2}}\right)\)
is a solution where \(\rho_{20}\) is a constant

\textbf{The argument on Differentiation}

If we assume that the Universe is filled with dust having then the
stress energy tensor \(T_{\mu \nu}\) becomes
\begin{displaymath}
\left(\begin{array}{cccc}
\rho & 0 & 0& 0 \\
 0 & -P& 0 & 0 \\
 0 & 0& -P & 0 \\
 0 & 0& 0 & -P
\end{array}\right)
\end{displaymath}
Then equation \ref {eq  rho} gets modified as
\begin{equation}\label{eq  rho pressure}
\frac{\dot{\rho}}{\rho+3P}=\frac{-3\dot{a}}{a}
\end{equation}
The equation of state determines P for each form of matter is
\begin{equation}
P=w \rho
\end{equation}
where w =-1 for dark energy ,\(w =\frac{2}{3}\) for dark matter
\footnote{is still to be confirmed through experiments},\(w
=\frac{2}{3}\) for normal matter and \(\frac{1}{3}\) for radiation.

Therefore for each kind of matter we have \(P_{i}=w_{i} \rho_{i}\)
Hence \(\rho_{i}+3P_{i}= \rho_{i}(1+w_{i})\).Hence each kind of
matter will distribute  differently due to different values of
\(\rho\) and P.Therefore in equation \ref{eq  rho pressure} the
value of
\begin{displaymath}
\rho+3P=\sum_{i}( \rho_{i}(1+w_{i})n_{i})
\end{displaymath}
where \(n_{i}\) is the number fraction of each type of matter.

Consider equation \ref{eq rho2}
\begin{displaymath}
 \dot{\rho_{2}}=
-\frac{3\dot{a_{1}}\rho_{2}}{a_{1}}+
\frac{3\dot{a_{1}}\rho_{1}}{a_{1}}\left(\frac{a_{2}}{a_{1}}\right)
-\frac{3\dot{a_{2}}\rho_{1}}{a_{1}}
\end{displaymath}
In the above equation \(\rho_{1}\) is the cumulative of all forms of
matter and \(a_{2}\) is also the cumulative of all forms of matter
since it is related to the space time structure and not on the kind
of matter.Therefore the only the matter dependent property is
\(\rho_{2}\).Therefore we say that matter does not differentiate
itself at the level of \(\rho_{1}\) But the differentiation occurs
at the level of \(\rho_{2}\) and \(\rho_{2}\) is a function of
position this means that different forms of matter occupy different
positions.

\(\rho_{2}\) is undifferentiated at WMAP stage and
\(\frac{\rho_{2}}{\rho}\) is of the order of \(10^{-5}\). Each
component is evolving differently by the time we get to galaxy
formation epoch.That means equation \ref{eq rho2} is now satisfied
independently for each component.Therefore the R.H.S of equation
\ref{eq rho2} is common but \(\rho_{2}\) is replaced by \((
\rho_{2_{i}}(1+w_{i})n_{i})\). where i runs from 1 to 4.i=1
represents dark energy;i=2 represents dark matter;i=3 represents
normal matter;i=4 represents radiation.The analogy of equation
\ref{eq rho2} with the diffusion equation \ref{eq diff rho}  with
the Gaussian solution require that each component has a diffusion
coefficient an d\(\sigma\) consistent with the observed data.Hence
clustering and differentiation has been explained.

\section{Two fluid clustering analog model of the Universe}

The Navier Stokes equation of fluid dynamics
\begin{displaymath}
\rho \left[\frac{\partial \vec{v}}{\partial
t}+(\vec{v}.\nabla)\vec{v}\right]=-\nabla P+ \eta
\nabla^{2}\vec{v}+\left(\zeta+\frac{\eta}{3} \right)grad (div
(\vec{v}) )
\end{displaymath}
Consider an incompressible fluid i.e, let \(div ( \vec{v} )=0 \)
Therefore the above equation becomes
\begin{displaymath}
\rho \left[\frac{\partial \vec{v}}{\partial
t}+(\vec{v}.\nabla)\vec{v}\right]=-\nabla P+ \eta \nabla^{2}\vec{v}
\end{displaymath}
Now at the scale of universe we say that dark energy is a fluid
which pushes dark matter and normal matter together to form super-
clusters of galaxies but once the structures are formed the dark
energy it pushes them apart and the super-clusters move away from
each other.One must zoom in and find similar process happening at
various scales.Assuming that the dark energy is a non-viscous fluid
in our model we put \(\eta=0\)
\begin{displaymath}
\rho \left[\frac{\partial \vec{v}}{\partial t}\right]=-\nabla P
\end{displaymath}
If \((\vec{v}.\nabla)\vec{v} << \nabla P\)
 The equation of state for dark energy is \(
P=-\rho\).Substituting this in the above equation we get
\begin{equation} \label{eq dark energy}
 \left[\frac{\partial \vec{v}}{\partial
t} \right]=\frac{\nabla \rho}{\rho}
\end{equation}
According to the equation of continuity
\begin{displaymath}
\frac{\partial \rho}{\partial t}+\nabla(\rho \vec{v})=0
\end{displaymath}
Now \(\rho(\nabla v)\)
\begin{displaymath}
\frac{\partial \rho}{\partial t}+\nabla(\rho)\vec{v}=0
\end{displaymath}
But from \ref{eq dark energy} we substitute \(\nabla(\rho)\) in the
above equation.Therefore
\begin{displaymath}
\frac{\partial \rho}{\partial t}=-\left(\rho\frac{\partial
\vec{v}}{\partial t}\right).\vec{v}
\end{displaymath}
\begin{displaymath}
\frac{\partial \rho}{\rho}=-\vec{v}.\vec{dv}
\end{displaymath}
(We have taken dark energy to be a fluid whose density is invariant
with position) Therefore
\begin{displaymath}
ln(\rho)=\frac{- v^{2}}{2}+b
\end{displaymath}

where \(\rho\)

 is the density of dark energy,\(\vec{v}\) the velocity
of dark energy and b is the constant of integration.
\newline
In case of dark matter the fluid equation is obtained as
follows.Consider the Navier-Stokes equation.The clustering of dark
and normal matter together suggests that \(\nabla P=0\) and the
formation of various kinds of galaxies like spiral suggests that the
viscous force is not negligible.But the fluid is assumed to be
incompressible i.e,  \(div ( \vec{v} )=0 \).Therefore the equation
for dark-matter normal matter mixture becomes
\begin{displaymath}
 \rho \left[\frac{\partial \vec{v}}{\partial
t}+(\vec{v}.\nabla)\vec{v}\right]=\eta \nabla^{2}\vec{v}
\end{displaymath}
The flow of any fluid depends through the boundary conditions,on
shape and dimensions of the body moving through the fluid and on its
velocity.Since the shape of the body is supposed given,its
geometrical properties are determined by one linear dimension,which
we denote by l.Let the velocity of the mainstream be u.Then any flow
is specified by three parameters, $\nu$,u and l where
\(nu=\frac{\eta}{\rho}\) is the kinematic viscosity.These quantities
have the following dimensions: \(\nu=\frac{cm^{2}}{sec} \ , \ l=cm \
, \  u=\frac{cm}{sec}\)\(\nu=\frac{cm^{2}}{sec} \ , \ l=cm \  , \
u=\frac{cm}{sec}\). It is easy to verify that only one dimensionless
quantity can be formed from the above three, namely
\(\frac{ul}{\nu}\).This combination is called the Reynolds number
and is denoted by R:
\begin{displaymath}
R=\frac{\rho u l}{\eta}=\frac{u l}{\nu}
\end{displaymath}

In case of the dark-matter,normal matter fluid one can assume the
flow to have small Reynolds number because of high value of
viscosity of fluid.Now the term \( (\vec{v}.\nabla)\vec{v})\)  is of
the order of magnitude \(\frac{u^{2}}{l}\).The quantity
\(\left(\frac{\eta}{\rho}\right)\nabla^{2}\vec{v}\) is of the order
of magnitude \(\frac{\eta u}{\rho l^{2}}\). The ratio of the two is
just the Reynolds number. Hence the term \(
((\vec{v}.\nabla)\vec{v})\) may be neglected if the Reynolds number
is small,and the equation of motion reduces to
\begin{displaymath}
 \rho \left[\frac{\partial \vec{v}}{\partial
t}\right]=\eta \nabla^{2}\vec{v}
\end{displaymath}
One can observe that this is nothing but the diffusion equation for
the velocity vector.The velocity vectors which are functions of
position and time satisfy the diffusion equation with diffusion
constant \(\frac{\eta}{\rho}\)

-

\section{Anisotropy and inhomogeneity of Density in universe}

    In standard models of Cosmology like Friedmann Robertson Walker ( FRW) models the universe is supposed to be
     homogenous and isotropic.But as we zoom in to smaller scales the inhomogeneity and anisotropy in the form of
     superclusters, clusters and galaxies appear. \\[10pt]

The formation of the \emph{Large Scale Structure} takes place due to
perturbation in the density of matter. Our aim is to introduce a
small perturbation in the density of the universe. This perturbation
is a function of $\theta$, $\phi$ and t.

The dynamic equations for the \emph{FRW-Model} have the form-
\begin{equation}
 \left(\frac{\dot{R}}{R}\right)^2 = \frac{8 \pi G\rho}{3}
\end{equation}
\begin{equation}
 2\frac{\ddot{R}}{R}+\left(\frac{\dot{R}}{R}\right)^2=-\frac{8 \pi G\rho}{3}
\end{equation}\\

The above two equation's are not independent of each other and from
the above we obtain the relation-
\begin{equation}
 \frac{\ddot{R}}{R}=-\left(\frac{\dot{R}}{R}\right)^2
\end{equation}\\
In equation (28) we put
 \[\rho \equiv \rho(t)+\delta\rho(\theta,\phi,t)\]
 and\\
 \[R\equiv R(t) +\delta R(\theta,\phi,t)\]\\
where $\delta R$ and $\delta \rho $ are perturbations of first order
in $R$ and $\rho$.

The equation (28) after solving leads to
\begin{equation}
2\left(\frac{\delta \dot R }{\dot{R}}-\frac{\delta
R}{R}\right)=\frac{\delta \rho}{\rho}
\end{equation}\\
In deriving the above equation the assumption is made that $R(t)$ and $\rho (t)$ follow Eq.. (28).
In the above Eq.. $R(t)$, $\dot{R}(t)$ and $\rho (t)$ can be derived from Eq. (28) as functions of time by making
the substitution $\rho=\frac{\rho_0 {R_0}^3}{{R}^3}$. Thus by knowing the variation of $\delta \rho$ with time an
ordinary differential equation governing the variation of $\delta R$ with time is obtained.\\[20pt]

In order to obtain the variation of $\delta \rho$ with time we make use of the equation which gives the variation of total density with time.\\
\begin{equation}
 \frac{\dot{\rho}}{\rho}=-3\frac{\dot{R}}{R}
\end{equation}\\

The above equation is derived by assuming that the total mass of the universe is constant.\\

Substituting $\frac{\dot{R}}{R}$ from Eq.. (28) in terms of $\rho$ leads to-\\
\begin{equation}
 \dot{\rho}=-3{\rho}^{3/2}\sqrt{\frac{8\pi G}{3}}
\end{equation}\\
 In which after substituting $\rho$=$\rho (t)+\delta\rho(\theta,\phi)$ and assuming that $\rho(t)$ follows Eq.. (31) we get\\
\begin{equation}
 \frac{\delta\dot{\rho}}{\delta\rho}=-\frac{9}{2}\sqrt{\frac{8\pi G \rho}{3}}
\end{equation}\\

From Eq.. (33) the time variation of $\delta\rho$ can be obtained
and thus the time variation of $\delta R$ can be obtained.

\section{Distribution and differentiation of various forms of matter
in the Universe}

Consider the differential equation of \(a_{2}\)v/s\(t\)
            \begin{displaymath}
                \ddot{a_{2}}- \frac{4a_{2}}{9 t^{2}}=0
            \end{displaymath}
            The solution of the above equation is
            \( \frac{c_{1}}{t^{\frac{1}{3}}}+c_{2} t^{\frac{4}{3}} \)
            where \(c_{1}\) and \(c_{2}\) are integration constants.
            One can see that this function attains an extremum at\(\frac{4 c_{2}}{c_{1}}\)

Supposing  the initial conditions were (i.e \(a_{2}=0 \  at \  t=0\)
we get
                the solution as
            \begin{equation}\label{sol a2}
            a_{2}=c(\theta , \phi)t^{\frac{4}{3}}
            \end{equation}

            Consider the differential equation \(\rho_{2}\)v/s \(t\)
            \begin{displaymath}
            \dot{\rho_{2}}=
            -\frac{3\dot{a_{1}}\rho_{2}}{a_{1}}+
            \frac{3\dot{a_{1}}\rho_{1}}{a_{1}}\left(\frac{a_{2}}{a_{1}}\right)
            -\frac{3\dot{a_{2}}\rho_{1}}{a_{1}}
            \end{displaymath}
            Substituting equation \ref{sol a2} in the above equation
            along with
            \begin{displaymath}
            \rho=\frac{\rho_{0}t_{0}^{2}}{t^{2}}
            \end{displaymath}
            and
            \begin{displaymath}
            a_{1}=a_{0}\left (\frac{t}{t_{0}} \right)^\frac{2}{3}
            \end{displaymath}
            Therefore
            \begin{displaymath}
            \dot{\rho_{2}}=
            -\frac{3\frac{d}{dt} \left({a_{0}\left (\frac{t}{t_{0}} \right)^\frac{2}{3}} \right)\rho_{2}}{a_{0}
            \left (\frac{t}{t_{0}} \right)^\frac{2}{3}}+
            \frac{3\frac{d}{dt} \left(a_{0}\left (\frac{t}{t_{0}} \right)^\frac{2}{3}\right)\frac{\rho_{0}t_{0}^{2}}{t^{2}}}{a_{0}
            \left (\frac{t}{t_{0}} \right)^\frac{2}{3}}\left(\frac{c(\theta , \phi )t^{\frac{4}{3}}}{a_{0}
            \left (\frac{t}{t_{0}} \right)^\frac{2}{3}}\right)
            -\frac{3\frac{d}{dt}\left(c(\theta , \phi )t^{\frac{4}{3}}\right)\frac{\rho_{0}t_{0}^{2}}{t^{2}}}{a_{0}
            \left (\frac{t}{t_{0}} \right)^\frac{2}{3}}
             \end{displaymath}
             \begin{displaymath}
            \dot{\rho_{2}}=
            -\frac{2 \rho_{2}}{t}+\frac{2c(\theta, \phi) \rho_{0}
            t_{0}^{\frac{8}{3}}}{a_{0}
t^{\frac{7}{3}}}-\frac{4c(\theta , \phi) \rho_{0}
t_{0}^{\frac{8}{3}}}{a_{0} t^{\frac{7}{3}} }
             \end{displaymath}
            We have \(t_{0}=13.7 \times 10^{9}\ years\ ,\
\rho_{0}=2.11 \times 10^{-29}\   kg/m^{3}\ \ \ and\ \ \ a_{0}=1\)
Hence
  \begin{displaymath}
            \dot{\rho_{2}}=\frac{2
            \rho_{2}}{t}+\frac{4.51 \times 10^{18}
c( \theta , \phi )}{t^{\frac{7}{3}}}
\end{displaymath}
\begin{displaymath}
        \rho_{2}=exp\left(-\int\frac{2}{t}dt
\right)\left[\int\left(\frac{4.51 \times 10^{18} c( \theta , \phi
)}{t^{\frac{7}{3}}}exp\left(\int\frac{2}{t}dt \right)\right)dt+K
\right]
\end{displaymath}
\begin{displaymath}
        \rho_{2}= t^{-2}\left[4.51 \times 10^{18} c( \theta , \phi
) \ \int\left(t^{2-\frac{7}{3} }\right)dt+K \right]
\end{displaymath}
\begin{displaymath}
        \rho_{2}=\left[\frac{4.51 \times 10^{18} c( \theta , \phi) \
t^{\frac{-4}{3}}}{2-\frac{4}{3}}+K t^{-2}  \right]
\end{displaymath}
\begin{displaymath}
\rho_{2}=\left[6.765 \times 10^{18} c( \theta , \phi) \
t^{\frac{-4}{3}}+K t^{-2}  \right]
\end{displaymath}
If we substitute the initial conditions at t=WMAP i.e
\(\frac{a_{2}}{a_{1}+a_{2}}=10^{-5}\) we get
\begin{displaymath}
 \frac{\frac{c_{1}}{t^{\frac{1}{3}}}+c_{2} t^{\frac{4}{3}}}{\frac{c_{1}}{t^{\frac{1}{3}}}+c_{2}
t^{\frac{4}{3}}+a_{0}\left (\frac{t}{t_{0}}
\right)^\frac{2}{3}}=10^{-5}
\end{displaymath}
\begin{equation}\label{linear eq a2}
2.00276 \times 10^{22} c_{2}+4.72702c_{1}=7.82709 \times 10^{8}
\end{equation}
However if we substitute \(a_{2}\) to be \(
\frac{c_{1}}{t^{\frac{1}{3}}}+c_{2} t^{\frac{4}{3}} \) in the
differential equation of \(\rho_{2}\) (if one is not sure about the
initial conditions of \(a_{2}\)) we get

 \begin{displaymath}
            \dot{\rho_{2}}=
            -\frac{3\frac{d}{dt} \left({a_{0}\left (\frac{t}{t_{0}} \right)^\frac{2}{3}} \right)\rho_{2}}{a_{0}
            \left (\frac{t}{t_{0}} \right)^\frac{2}{3}}+
            \frac{3\frac{d}{dt} \left(a_{0}\left (\frac{t}{t_{0}} \right)^\frac{2}{3}\right)\frac{\rho_{0}t_{0}^{2}}{t^{2}}}{a_{0}
            \left (\frac{t}{t_{0}} \right)^\frac{2}{3}}\left(\frac{\frac{c_{1}}{t^{\frac{1}{3}}}+c_{2}t^{\frac{4}{3}}}{a_{0}
            \left (\frac{t}{t_{0}} \right)^\frac{2}{3}}\right)
            -\frac{3\frac{d}{dt}\left(\frac{c_{1}}{t^{\frac{1}{3}}}+c_{2}t^{\frac{4}{3}}\right)\frac{\rho_{0}t_{0}^{2}}{t^{2}}}{a_{0}
            \left (\frac{t}{t_{0}} \right)^\frac{2}{3}}
             \end{displaymath}
  \begin{displaymath}
            \dot{\rho_{2}}=
            -\frac{2 \rho_{2}}{t}+\frac{2c_{1}\rho_{0}
            t_{0}^{\frac{8}{3}}}{a_{0}
t^{4}}+\frac{c_{1} \rho_{0} t_{0}^{\frac{8}{3}}}{a_{0} t^{4}
}+\frac{2c_{2} \rho_{0}
            t_{0}^{\frac{8}{3}}}{a_{0}
t^{\frac{7}{3}}}-\frac{4c_{2} \rho_{0} t_{0}^{\frac{8}{3}}}{a_{0}
t^{\frac{7}{3}} }
 \end{displaymath}
 We have \(t_{0}=13.7 \times 10^{9}\ years\ ,\
\rho_{0}=2.11 \times 10^{-29}\   kg/m^{3}\ \ \ and\ \ \ a_{0}=1\)
Hence
\begin{displaymath}
            \dot{\rho_{2}}=\frac{-2
            \rho_{2}}{t}+\frac{6.76502 \times 10^{18}
c_{1}}{t^{4}}+\frac{(-4.51001\times 10^{18} c_{2})}{t^{\frac{7}{3}}}
\end{displaymath}
\begin{displaymath}
            \dot{\rho_{2}}+\frac{2
            \rho_{2}}{t}=\frac{6.76502  \times 10^{18}
c_{1}}{t^{4}}+\frac{(-4.51001 \times 10^{18}
c_{2})}{t^{\frac{7}{3}}}
\end{displaymath}

Call \(2\) as \(k_{1}\),\(6.76502 \times 10^{18} c_{1}\) as
\(k_{2}\) and\(-4.51001 \times 10^{18} c_{2}\) as \(k_{3}\).
Therefore the above equation becomes
\begin{displaymath}
\dot{\rho_{2}}+\frac{k_{1}}{t}=\frac{k_{2}}{t^{4}}+\frac{k_{3}}{t^{\frac{7}{3}}}
\end{displaymath}
The above equation is a linear first order differential equation
whose solution is
\begin{displaymath}
\rho_{2}=\left(\frac{k_{2} t^{-3}}{k_{1}-3}\right)+\left(
\frac{k_{3}t^\frac{-4}{3}}{k_{1}-\frac{4}{3}}\right)+kt^{-k_{1}}
\end{displaymath}
where k is the constant of integration.Substituting the values of
\(k_{1},k_{2}\) and \(k_{3}\) respectively we get
\begin{displaymath}
\rho_{2}=\left(-6.76502 \times 10^{18} c_{1} t^{-3}\right)+\left(
-6.765 \times 10^{18} c_{2}t^\frac{-4}{3}\right)+kt^{-2}
\end{displaymath}
where k is the constant of integration.
\newline
We know that at \(t=WMAP=300,000\) years
\begin{equation}\label{eq wmap}
\frac{\rho_{2}}{\rho_{1}+\rho_{2}}=10^{-5}
\end{equation}
\(\rho_{2}\) at \(t=300,000\) years is calculated to be \(-7.9724
\times 10^{-21} c_{1} -33.778 c_{2}+1.116 \times 10^{-26}\) and
\(\rho_{1}\) at \(t=300,000\) years is equal to \(4.40029 \times
10^{-20}\). Therefore at t=WMAP substituting \(\rho_{1}\) and
\(\rho_{2}\) in \ref{eq wmap} we get
\begin{displaymath}
3.3778 \times 10^{6}c_{2}-7.9724\times 10^{-16}c_{1}+1.116 \times
10^{-26}= 4.40029 \times 10^{-20}
\end{displaymath}
Now $c_{1}$ and $c_{2}$ are actually functions of \(\theta\) and
\(\phi\).Therefore imagine \(c_{1} =\alpha f( \theta,\phi)\) and
\(c_{2}=\beta g(\theta, \phi)\))

\section{Calculation of Einstein tensor for the modified Friedmann Robertson
Walker metric}
 Let us call a[t] as R[t]. The modified Robertson
Walker metric is given by
\begin{displaymath}
(ds)^{2}=(dt)^{2}-[R(t)+b(t,\theta,\phi)]^{2}(dr)
^{2}-[R(t)+b(t,\theta,\phi)]^{2}(r) ^{2}(d \theta) ^{2}-
[R(t)+b(t,\theta,\phi)]^{2}(r) ^{2}sin^{2}(\theta)(d \phi) ^{2}
\end{displaymath}
where $b(t,\theta,\phi)$ is the perturbation introduced in $R(t)$
Then the components of the Einstein Tensor for the modified
Robertson Walker metric which introduces inhomogeneity is

Below R is a function of t and b is a function of  \(t\ , \ \theta \
, \ \phi\).Note that for each of the terms below the term in the
first bracket is nothing but the terms of the Einstein tensor for
the Robertson-Walker metric.Also in evaluating the Einstein tensor a
binomial expansion was made and terms involving higher powers of b
or the derivatives of  b have been neglected. The complete Einstein
tensor was evaluated using Mathematica.

\section{Einstein tensor for perturbed FRW metric}

\begin{displaymath}
G_{00}=\frac{-1}{r^{2}R^{4}}
 \left[
\begin{array}{c} \left[[3(rRR^{'})^{2} \right]+b
\left( 6R(rR^{'})^{2}- 12(rR^{'})^{2}R \right) +\frac{\partial
b}{\partial t} (6(rR)^{2}R^{'})\\+\frac{\partial b}{\partial
\theta}(-2R cot(\theta))+\frac{\partial^{2}b}{\partial
\theta^{2}}(-2R)+\frac{\partial^{2}b}{\partial \phi^{2}}(-2R
cosec^{2}(\theta))
\end{array} \right]
\end{displaymath}
\begin{displaymath}
G_{01}=0
\end{displaymath}
\begin{displaymath}
G_{02}=\frac{2}{R^{2}}\left[-R^{'}\left(\frac{\partial b}{\partial
\theta}\right)+R \left( \frac{\partial^{2} b}{\partial t
\partial \theta}\right)\right]
\end{displaymath}
\begin{displaymath}
G_{03}=\frac{2}{R^{2}}\left[-R^{'}\left(\frac{\partial b}{\partial
\phi}\right)+R \left( \frac{\partial^{2} b}{\partial t
\partial \phi}\right)\right]
\end{displaymath}
\begin{displaymath}
G_{10}=0
\end{displaymath}
\begin{displaymath}
G_{11}=\frac{1}{(rR)^{2}}
 \left[
\begin{array}{c}
\left[(rRR^{'})^{2}+2r^{2}R^{3}R^{''}\right]
+b \left(2R(rR^{'})^{2}+6(rR)^{2}R^{''}+\frac{-2}{R}\left((rRR^{'})^{2}+2r^{2}R^{3}R^{''}\right)\right)\\
+\frac{\partial b}{\partial t}(2(rR)^{2}R^{'})+\frac{\partial
b}{\partial \theta}(-R cot(\theta)) +\frac{\partial^{2} b}{\partial
t^{2}}(2r^{2}R^{3}) +\frac{\partial^{2} b}{\partial \theta^{2}}(-R)+
\frac{\partial^{2} b}{\partial \phi^{2}}(-R cosec^{2}\theta )
\end{array} \right]
\end{displaymath}
\begin{displaymath}
G_{12}=\frac{-1}{rR}\left(\frac{\partial b}{\partial \theta}\right)
\end{displaymath}
\begin{displaymath}
G_{13}=\frac{-1}{rR}\left(\frac{\partial b}{\partial \phi}\right)
\end{displaymath}
\begin{displaymath}
G_{20}=\frac{2}{R^{2}}\left[-R^{'}\left(\frac{\partial b}{\partial
\theta}\right)+R \left( \frac{\partial^{2} b}{\partial t
\partial \theta}\right)\right]
\end{displaymath}
\begin{displaymath}
G_{21}=\frac{-1}{rR}\left(\frac{\partial b}{\partial \theta}\right)
\end{displaymath}
\begin{displaymath}
G_{22}=\frac{1}{(rR)^{2}}
 \left[
\begin{array}{c}
\left[(rRR^{'})^{2}+2r^{2}R^{3}R^{''}\right]
+b \left(2R(rR^{'})^{2}+6(rR)^{2}R^{''}+\frac{-2}{R}\left((rRR^{'})^{2}+2r^{2}R^{3}R^{''}\right)\right)\\
+\frac{\partial b}{\partial t}(2(rR)^{2}R^{'})+\frac{\partial
b}{\partial \theta}(-R cot(\theta)) +\frac{\partial^{2} b}{\partial
t^{2}}(2r^{2}R^{3}) + \frac{\partial^{2} b}{\partial \phi^{2}}(-R
cosec^{2}\theta )
\end{array} \right]
\end{displaymath}
\begin{displaymath}
G_{23}= \frac{1}{R}\left[\left(\frac{\partial b}{\partial
\phi}\right)(-cot\theta)+\frac{\partial^{2} b}{\partial \theta
\partial \phi}\right]
\end{displaymath}
\begin{displaymath}
G_{30}=\frac{2}{R^{2}}\left[-R^{'}\left(\frac{\partial b}{\partial
\phi}\right)+R \left( \frac{\partial^{2} b}{\partial t
\partial \phi}\right)\right]
\end{displaymath}
\begin{displaymath}
G_{31}=\frac{-1}{rR}\left(\frac{\partial b}{\partial \phi}\right)
\end{displaymath}
\begin{displaymath}
G_{32}= \frac{1}{R}\left[\left(\frac{\partial b}{\partial
\phi}\right)(-cot\theta)+\frac{\partial^{2} b}{\partial \theta
\partial \phi}\right]
\end{displaymath}
\begin{displaymath}
G_{33}=\frac{1}{R^{2}} \left[
\begin{array}{c}
\left[\left( rRR^{'} sin \theta \right)^{2}+2r^{2}R^{3}sin^{2}\theta
R^{''} \right]\\+b\left(2R(rR^{'}sin \theta)^{2}+6R{''}(rR sin
\theta)^{2} -\frac{2}{R}\left( rRR^{'} sin \theta
\right)^{2}+2r^{2}R^{3}sin^{2}\theta R^{''} \right)
\\ +\frac{\partial b}{\partial t}(2R(rR^{'}sin
\theta)^{2})+\frac{\partial^{2} b}{\partial
t^{2}}(2r^{2}R^{3}sin^{2}\theta)\end{array} \right]
\end{displaymath}
If we consider the standard stress-energy tensor (which is dust with
pressure) in which the off diagonal terms are zero then
\(b(t,\theta,\phi)=0\) which removes the inhomogeneity which is
wanted in the first place.So one must think of an appropriate stress
energy tensor which when equated to the Einstein tensor gives a
solution for \(b(t,\theta,\phi)\)

The off diagonal terms in Einstein's equation are set to zero in
this approximation  as case (1), and in the right hand side, the off
diagonal elements of the  stress energy tensor are set to zero.
Consider only the diagonal terms.

\section{Approximate solution}
\begin{displaymath}
R=a_{1}=a_{0}\left (\frac{t}{t_{0}} \right)^\frac{2}{3}
\end{displaymath}
Now since \(a_{0}=1\)
\begin{displaymath}
\Longrightarrow R^{'}=\frac{2}{3t_{0}}\left (\frac{t}{t_{0}}
\right)^\frac{-1}{3}
\end{displaymath}
and
\begin{displaymath}
\Longrightarrow R^{''}=\frac{-2}{9t_{0}^{2}}\left (\frac{t}{t_{0}}
\right)^\frac{-4}{3}
\end{displaymath}
Also
\begin{displaymath}
b=a_{2}=c(\theta , \phi)t^{\frac{4}{3}}
\end{displaymath}
\begin{displaymath}
\frac{\partial b}{\partial t}=\frac{4}{3}c(\theta ,
\phi)t^{\frac{1}{3}}
\end{displaymath}
\begin{displaymath}
\frac{\partial^{2} b}{\partial t^{2}}=\frac{4}{9}c(\theta ,
\phi)t^{\frac{-2}{3}}
\end{displaymath}
Also
\begin{displaymath}
\frac{\partial b}{\partial \theta }=\frac{\partial c(\theta , \phi)
}{\partial \theta}t^{\frac{4}{3}}
\end{displaymath}
Neglecting the higher derivatives of \(\theta\) and \(\phi\) in
\(G_{00}\) we get
\begin{displaymath}
G_{00}=\frac{-1}{r^{2}R^{4}}
 \left[
\begin{array}{c} \left(3(rRR^{'})^{2} \right)+b
\left( -6(rR^{'})^{2}R \right) +\frac{\partial
b}{\partial t} (6(rR)^{2}R^{'})\\+\frac{\partial b}{\partial
\theta}(-2R cot(\theta))
\end{array} \right]
\end{displaymath}
\begin{displaymath}
G_{00}=\frac{-1}{r^{2}\left[\left (\frac{t}{t_{0}}
\right)^\frac{2}{3}\right]^{4}}
 \left[
\begin{array}{c} \left(3\left(r\left (\frac{t}{t_{0}}
\right)^\frac{2}{3}\frac{2}{3t_{0}}\left (\frac{t}{t_{0}}
\right)^\frac{-1}{3}\right)^{2} \right)+c(\theta , \phi)t^{\frac{4}{3}}
\left( -6\left(\frac{2r}{3t_{0}}\left (\frac{t}{t_{0}}
\right)^\frac{-1}{3} \right)^{2}\left (\frac{t}{t_{0}} \right)^\frac{2}{3} \right)\\ +\frac{4c(\theta ,
\phi)t^{\frac{1}{3}}}{3}\left(6(r\left (\frac{t}{t_{0}} \right)^\frac{2}{3})^{2}\frac{2}{3t_{0}}\left (\frac{t}{t_{0}}
\right)^\frac{-1}{3}\right)+\frac{\partial c(\theta , \phi)
}{\partial \theta}t^{\frac{4}{3}}\left(-2 \left(\frac{t}{t_{0}} \right)^\frac{2}{3} cot(\theta)\right)
\end{array} \right]
\end{displaymath}
Therefore \(G_{00}\) simplifies to give
\begin{displaymath}
\frac{-1}{r^{2}}\left(\frac{t}{t_{0}} \right)^{\frac{8}{3}} \left[
\frac{4 r^{2}}{3 t_{0}^{2}}\left(
\frac{t}{t_{0}}\right)^\frac{2}{3}+
\frac{8cr^{2}t^{\frac{4}{3}}}{3t_{0}^{2}}-2cot(\theta)t^{\frac{4}{3}}\left(\frac{t}{t_{0}}
\right)^\frac{2}{3}\frac{\partial c(\theta , \phi) }{\partial
\theta}
 \right]
\end{displaymath}
Neglecting the higher derivatives of \(\theta\) and \(\phi\) in
\(G_{11}\) we get
\begin{displaymath}
G_{11}=\frac{1}{(rR)^{2}}
 \left[
\begin{array}{c}
\left[(rRR^{'})^{2}+2r^{2}R^{3}R^{''}\right]
+b \left(2R(rR^{'})^{2}+6(rR)^{2}R^{''}+\frac{-2}{R}\left((rRR^{'})^{2}+2r^{2}R^{3}R^{''}\right)\right)\\
+\frac{\partial b}{\partial t}(2(rR)^{2}R^{'})+\frac{\partial
b}{\partial \theta}(-R cot(\theta)) +\frac{\partial^{2} b}{\partial
t^{2}}(2r^{2}R^{3})
\end{array} \right]
\end{displaymath}
\begin{displaymath}
G_{11}=\frac{1}{\left(r\left (\frac{t}{t_{0}}
\right)^\frac{2}{3}\right)^{2}}
 \left[
\begin{array}{c}
\left[\left(r\left(\frac{t}{t_{0}} \right)^\frac{2}{3}\frac{2}{3t_{0}}\left(\frac{t}{t_{0}}
\right)^\frac{-1}{3}\right)^{2}+2r^{2}\left[\left(\frac{t}{t_{0}} \right)^\frac{2}{3}\right]^{3}\frac{-2}{9t_{0}^{2}}\left (\frac{t}{t_{0}}
\right)^\frac{-4}{3}\right] \\
+c(\theta , \phi)t^{\frac{4}{3}} \left(2\left(\frac{t}{t_{0}} \right)^\frac{2}{3}\left(\frac{2r}{3t_{0}}\left(\frac{t}{t_{0}}
\right)\right)^{2}
+6\left(r\left(\frac{t}{t_{0}} \right)^\frac{2}{3}\right)^{2}\frac{-2}{9t_{0}^{2}}\left (\frac{t}{t_{0}}\right)^\frac{-4}{3}\right)\\
+c(\theta , \phi)t^{\frac{4}{3}}\left(\frac{-2}{\left (\frac{t}{t_{0}} \right)^\frac{2}{3}}\left(\left(r\left
(\frac{t}{t_{0}} \right)^\frac{2}{3}\frac{2}{3t_{0}}\left
(\frac{t}{t_{0}}
\right)^\frac{-1}{3}\right)^{2}+2r^{2}\left[\left(\frac{t}{t_{0}} \right)^\frac{2}{3}\right]^{3}\frac{-2}{9t_{0}^{2}}\left (\frac{t}{t_{0}}
\right)^\frac{-4}{3}\right)\right)\\
+\frac{4}{3}c(\theta,\phi)t^{\frac{1}{3}}\left(2(r\left (\frac{t}{t_{0}}\right)^\frac{2}{3})^{2}\frac{2}{3t_{0}}\left (\frac{t}{t_{0}}
\right)^\frac{-1}{3}\right)\\
-\frac{\partial c(\theta , \phi)}{\partial \theta}t^{\frac{4}{3}}\left(\left(\frac{t}{t_{0}} \right)^\frac{2}{3} cot(\theta)\right)\\
+\frac{4}{9}c(\theta ,\phi)t^{\frac{-2}{3}}\left(2r^{2}\left[\left(\frac{t}{t_{0}} \right)^ \frac{2}{3}\right]^{3}\right)
\end{array} \right]
\end{displaymath}
\begin{displaymath}
G_{11}=\frac{1}{\left(r\left (\frac{t}{t_{0}}
\right)^\frac{2}{3}\right)^{2}}
\left[\frac{28cr^{2}t^{\frac{4}{3}}}{9t_{0}^{2}} -\frac{\partial
c(\theta , \phi)}{\partial \theta}t^{\frac{4}{3}}\left(\left
(\frac{t}{t_{0}} \right)^\frac{2}{3}
cot(\theta)\right)+\frac{8cr^{2}t^{\frac{-2}{3}}}{9}\left
(\frac{t}{t_{0}} \right)^{2} \right]
\end{displaymath}
Observation yields
\begin{displaymath}
G_{11}+\frac{R}{\left(r\left (\frac{t}{t_{0}}
\right)^\frac{2}{3}\right)^{2}}\left[\frac{\partial^{2} b}{\partial
\theta^{2}}\right]=G_{22}
\end{displaymath}
Neglect the higher derivatives of \(b=c(\theta,\phi)
t^{\frac{4}{3}}\)yields
\begin{displaymath}
G_{11}=G_{22}
\end{displaymath}
\(G_{33}\)does not yield a differential equation for \(b(\theta,
\phi)\) and hence it is redundant for this approximation.

Therefore
\begin{displaymath}
G_{00}=\frac{-1}{r^{2}}\left(\frac{t}{t_{0}} \right)^{\frac{-8}{3}}
\left[ \frac{4 r^{2}}{3 t_{0}^{2}}\left(
\frac{t}{t_{0}}\right)^\frac{2}{3}+
\frac{8cr^{2}t^{\frac{4}{3}}}{3t_{0}^{2}}-2cot(\theta)t^{\frac{4}{3}}\left(\frac{t}{t_{0}}
\right)^\frac{2}{3}\frac{\partial c(\theta , \phi) }{\partial
\theta}
 \right]
\end{displaymath}
\begin{displaymath}
G_{11}=G_{22}=\frac{1}{\left(r\left (\frac{t}{t_{0}}
\right)^\frac{2}{3}\right)^{2}}
\left[\frac{28ct^{\frac{4}{3}r^{2}}}{9t_{0}^{2}} -\frac{\partial
c(\theta , \phi)}{\partial \theta}t^{\frac{4}{3}}\left(\left
(\frac{t}{t_{0}} \right)^\frac{2}{3}
cot(\theta)\right)+\frac{8cr^{2}t^{\frac{-2}{3}}}{9}\left
(\frac{t}{t_{0}} \right)^{2} \right]
\end{displaymath}
Substitute for t the WMAP time which is \(3\times 10^{5}\times
365\times 24 \times 60 \times 60\) and for \(t_{0}\) \(13.6 \times
10^{9}\times 365\times 24 \times 60 \times 60\) seconds. These in
the units \(c=1\) will be measured in meters i.e
\begin{displaymath}
t_{WMAP}=3\times 10^{5}\times 365\times 24 \times 60 \times 60
\times 3 \times 10^{8} \ meters=2.84\times 10^{21} \ meters
\end{displaymath}
\begin{displaymath}
t_{0}=13.6 \times 10^{9}\times 365\times 24 \times 60 \times 60
\times 3 \times 10^{8} \ meters=1.29\times 10^{26} \ meters
\end{displaymath}
The first term in \(G_{00}\) is far less in magnitude than the other
terms. Hence
\begin{displaymath}
G_{00}=\frac{-1}{r^{2}}\left(\frac{t}{t_{0}} \right)^{\frac{-8}{3}}
\left[\frac{8cr^{2}t^{\frac{4}{3}}}{3t_{0}^{2}}-2cot(\theta)t^{\frac{4}{3}}\left(\frac{t}{t_{0}}
\right)^\frac{2}{3}\frac{\partial c(\theta , \phi) }{\partial
\theta} \right]
\end{displaymath}
\begin{displaymath}
G_{11}=G_{22}=\frac{-1}{r^{2}}\left(\frac{t}{t_{0}}
\right)^{\frac{-8}{3}}
\left[\frac{4cr^{2}t^{\frac{4}{3}}}{t_{0}^{2}}-cot(\theta)t^{\frac{4}{3}}\left(\frac{t}{t_{0}}
\right)^\frac{2}{3}\frac{\partial c(\theta , \phi) }{\partial
\theta} \right]
\end{displaymath}
Now in \(c=1\) units \(G_{00}=8 \pi G (\rho_{1}+\rho_{2})\). This
includes the first and second order terms. The \(G_{00}\) which we
obtained is  second order, and approximating  \(\rho_{2}\approx 0\).

  \(8 \pi G \rho_{2}\) has a very small value compared to terms on
the L.H.S.
\section{case 1}
\begin{displaymath}
\frac{-1}{r^{2}}\left(\frac{t}{t_{0}} \right)^{\frac{-8}{3}}
\left[\frac{8cr^{2}t^{\frac{4}{3}}}{3t_{0}^{2}}-2cot(\theta)t^{\frac{4}{3}}\left(\frac{t}{t_{0}}
\right)^\frac{2}{3}\frac{\partial c(\theta , \phi) }{\partial
\theta} \right]=0
\end{displaymath}
\begin{displaymath}
\Longrightarrow
\left[\frac{8cr^{2}t^{\frac{4}{3}}}{3t_{0}^{2}}-2cot(\theta)t^{\frac{4}{3}}\left(\frac{t}{t_{0}}
\right)^\frac{2}{3}\frac{\partial c(\theta , \phi) }{\partial
\theta} \right]=0
\end{displaymath}
\begin{displaymath}
\Longrightarrow \frac{4 r^{2}
\left(t_{WMAP}\right)^{\frac{-2}{3}}\left(t_{0}\right)^{\frac{-4}{3}}}{3}\left[tan(\theta)\,\partial
\theta\right]=\frac{\partial\left[ c(\theta , \phi)\right]}{c(\theta
, \phi)}
\end{displaymath}
\begin{displaymath}
\Longrightarrow \frac{4 r^{2}
\left(t_{WMAP}\right)^{\frac{-2}{3}}\left(t_{0}\right)^{\frac{-4}{3}}}{3}\int
tan(\theta)\,\partial \theta=\int\frac{\partial \left[ c(\theta ,
\phi)\right]}{c(\theta , \phi)}
\end{displaymath}
\begin{displaymath}
\Longrightarrow 1.024 \times 10^{-49}r^{2}ln(sec
(\theta))+ln(g(\phi))=ln(c(\theta , \phi))
\end{displaymath}
where \(g(\phi)\) is the constant of integration.
\begin{displaymath}
\Longrightarrow c(\theta , \phi)= g(\phi)[sec \theta]^{1.024 \times
10^{-49}r^{2}}
\end{displaymath}
Now in \(c=1\) units \(G_{11}=G_{22}=8 \pi G (P_{1}+P_{2})\). Now
the \(G_{11}\) and \(G_{22}\) which we obtained is called second
order and approximating i.e \(P_{2}\approx 0\) then \(8 \pi G
P_{2}\) has a very small value compared to terms on the L.H.S. Hence
we get
\begin{displaymath}
\frac{-1}{r^{2}}\left(\frac{t}{t_{0}} \right)^{\frac{-8}{3}}
\left[\frac{4cr^{2}t^{\frac{4}{3}}}{t_{0}^{2}}-cot(\theta)t^{\frac{4}{3}}\left(\frac{t}{t_{0}}
\right)^\frac{2}{3}\frac{\partial c(\theta , \phi) }{\partial
\theta} \right]=0
\end{displaymath}
\begin{equation}\label{eq approx}
\Longrightarrow 4 r^{2}
\left(t_{WMAP}\right)^{\frac{-2}{3}}\left(t_{0}\right)^{\frac{-4}{3}}\left[tan(\theta)\,\partial
\theta\right]=\frac{\partial\left[ c(\theta , \phi)\right]}{c(\theta
, \phi)}
\end{equation}
\begin{displaymath}
\Longrightarrow 4 r^{2}
\left(t_{WMAP}\right)^{\frac{-2}{3}}\left(t_{0}\right)^{\frac{-4}{3}}\int
tan(\theta)\,\partial \theta=\int\frac{\partial \left[ c(\theta ,
\phi)\right]}{c(\theta , \phi)}
\end{displaymath}
\begin{displaymath}
\Longrightarrow 3\times 1.024 \times 10^{-49}r^{2}ln(sec
(\theta))+ln(h(\phi))=ln(c(\theta , \phi))
\end{displaymath}
where \(h(\phi)\) is the constant of integration.
\begin{displaymath}
\Longrightarrow c(\theta , \phi)= h(\phi)[sec \theta]^{3 \times
1.024 \times 10^{-49}r^{2}}
\end{displaymath}
Now for \(r=r_{WMAP}=r_{0}\left(\frac{t_{WMAP}}{t_{0}}
\right)^{\frac{2}{3}}=7.139 \times 10^{22}\). Finally
\begin{displaymath}
\Longrightarrow c(\theta , \phi)= h(\phi)[sec \theta]^{4.68 \times
10^{-3}}
\end{displaymath}
Using the \(G_{11}\) equation one gets
\begin{displaymath}
c(\theta , \phi)=g(\phi)[sec \theta]^{1.56 \times 10^{-3}}
\end{displaymath}
Therefore
\begin{displaymath}
c(\theta , \phi)=g(\phi)[sec \theta]^{1.56 \times
10^{-3}}=h(\phi)[sec \theta]^{4.68 \times 10^{-3}}
\end{displaymath}
Now in obtaining \(c(\theta,\phi)\) we have taken
\(\frac{\partial^{2}b}{\partial \theta^{2}} \approx 0\) and
neglected this term. So this expression is valid for those values of
\(\theta\) where \(\frac{\partial^{2}c(\theta , \phi)}{\partial
\theta^{2}} \approx 0\) i.e \(\frac{\partial c(\theta , \phi)
}{\partial \theta} \approx\) constant and any deviation will be
linear in \(\theta\). Inserting the solution we find that the first
derivative is proportional approximately to \(tan(\theta\))\ equals
\(\theta\) for small \(\theta\). This happens for those values of
\(\theta\) for which \(c(\theta , \phi)tan(\theta)\approx\)
constant, which is true for values of \(\theta\) close to \(0\).
Hence the given function describes anisotropy the best for values of
\(\theta\) around \(0\).

To evaluate the functions of \(\phi\) one must need another equation
in \(c(\theta , \phi)\) v/s \(\phi\) for which one must take higher
order perturbations. But as of now \(b(t,\theta,\phi)=g(\phi)[sec
\theta]^{1.56 \times 10^{-3}}t^{\frac{-4}{3}}=h(\phi)[sec
\theta]^{4.68 \times 10^{-3}}t^{\frac{-4}{3}}\).  The off diagonal
terms of \(G_{ij}\) were set to zero. But since we now have the
expression for \(b(t,\theta,\phi)\) we can now substitute it in the
off diagonal terms say \(G_{12}\) and \(G_{21}\) and since \(G_{12}=
8 \pi G T_{12}\) and \(G_{21}= 8 \pi G T_{21}\) we have now actually
got two terms for our stress energy tensor. i.e
\begin{displaymath}
T_{12}=T_{21}=\frac{-g(\phi)(\alpha sec^{\alpha}\theta
tan\theta)t_{wmap}^{\frac{4}{3}}}{r_{wmap}}\left[\frac{t_{wmap}}{t_{0}}\right]^{-\frac{2}{3}}
\end{displaymath}

where \(\alpha=1.56 \times 10^{-3}\) . Similarly the other
components of the stress energy tensor can be determined. They would
have derivatives of \(h(\phi)\).  \(T_{12} \) and \(T_{21}\) involve
\(h(\phi)\). But it  can be determined by not neglecting
\(\frac{\partial^{2}b}{\partial \phi^{2}}\). Thus the stress energy
tensor can be completely determined and seen to correspond to
inhomogeneity. This can be interpreted as introducing perturbations
in the ideal fluid contained in Universe to obtain the non-ideal
case.

\section{case 2}
 In the above calculation we have  \({\rho_{1}}\) non zero for
first order and \(\rho_{2}=0\). But \(\rho_{2} = G_{00}\) for second
order. Here
\begin{displaymath}
\rho=\frac{\rho_{0}t_{0}^{2}}{t^{2}}
\end{displaymath}
\begin{displaymath}
\left[6.765 \times 10^{18}\times(3\times10^{8})^{\frac{8}{3}} c(
\theta , \phi) \ t^{\frac{-4}{3}}+K t^{-2}  \right]
\end{displaymath}
where \(\rho_{0}=2.11 \times 10^{-29}\) and K is a constant
\(t=t_{wmap}\) and \(t_{0}=present  time\)
\begin{displaymath}
G_{00}=\frac{-1}{r^{2}}\left(\frac{t}{t_{0}} \right)^{\frac{8}{3}}
\left[ \frac{4 r^{2}}{3 t_{0}^{2}}\left(
\frac{t}{t_{0}}\right)^\frac{2}{3}+
\frac{8cr^{2}t^{\frac{4}{3}}}{3t_{0}^{2}}-2cot(\theta)t^{\frac{4}{3}}\left(\frac{t}{t_{0}}
\right)^\frac{2}{3}\frac{\partial c(\theta , \phi) }{\partial
\theta}
 \right]
\end{displaymath}
\begin{displaymath}
 =\rho_{1}+\rho_{2}=\frac{\rho_{0}t_{0}^{2}}{t^{2}}=\left[6.765 \times 10^{18}\times(3\times10^{8})^{\frac{8}{3}} c(
\theta , \phi) \ t^{\frac{-4}{3}}+K t^{-2}  \right]
\end{displaymath}
which yields
\begin{displaymath}
(cot(\theta) \times 10^{8})\frac{\partial c(\theta , \phi)
}{\partial \theta}- c(\theta , \phi)(69 \times 10^{12})+K
t^{-2}+4.35\times10^{-20}=0
\end{displaymath} This equation can again be solved for \(c(\theta,\phi)\) , and hence
for b and the complete Einstein tensor and stress energy tensor can
be determined. The model is solved at WMAP time.

\section{Conclusion}

The correlation between the 2nd order gravity wave analysis which
gives the space time 'wrinkles' on the FRW background and the
density and curvature perturbations of second order that give the
matter distribution on FRW background is expected, as the Einstein
equations remain valid order by order. The correlation of the
observed WMAP radiation data to the theoretical model is also
suggested. The WMAP data has been explained by other models such as
inflation driven inhomogeneity. In this work the contribution of
gravity waves , non linearly coupled on the FRW background,
organizing  the density and curvature inhomogeneity was
investigated.

\section{ACKNOWLEDGEMENT}

We thank the NIUS program and its staff as well as the HBCSE for the
facilities provided.

\section{References}
\begin{itemize}
\item
An Introduction to Cosmology-Jayant Vishnu Narlikar
\item
Cosmological Physics-John A.Peacock
\item
Gravitation and Cosmology-Steven Weinberg
\item
WMAP 3 year results:Implications for Cosmology-D.N.Spergel and
others
\item
From arxiv.org:
\item
Cosmology at the Millennium-by Michael Turner (astro-ph/9901113)
\item
Dark matter in groups and clusters of
galaxies.(astro-ph/9909437)
\item
astro-ph/0304483-Gibson;Schild
\item
astro-ph/0609686-J.Einasto
\item
astro-ph/0604539-J.Einasto et.al
\item Mathematica for Physicists
\item Fluid Mechanics by L.D.Landau and E.M.Lifshitz
\item Relativity Demystified by David MacMohan
\item Cosmology and Astrophysics through problems by T.Padmanabhan
\end{itemize}

\end{document}